\documentclass[usenatbib]{mn2e}
\usepackage{natbib}
\usepackage{apjfonts}
\usepackage{amsmath}

\usepackage{epsfig}
\usepackage{graphicx}
\usepackage{amssymb}
\usepackage{aas_macros}
\usepackage{color}

\newcommand{\oversim}[2]{\protect{\mbox{\lower0.5ex\vbox{%
   \baselineskip=0pt\lineskip=0.2ex
   \ialign{$\mathsurround=0pt #1\hfil##\hfil$\crcr#2\crcr\sim\crcr}}}}} 

\catcode`\"=\active\let"=\"  

\def\3{{\ss} }

\def\c12{{1\over 2}}

\def\d{{\rm d}}   
   
\def\plusplus{\raise 0.3ex\hbox{${\scriptstyle ++}$}{}}

\bibliography{biblio}

\begin{document}   
%\title[Evolution of dynamic tracers in time-dependent potentials]{Evolution of dynamic tracers in time-dependent potentials}
%\title[Non-equilibrium dynamics of collisionless systems]{Non-equilibrium dynamics of collisionless systems in time-dependent gravitational potentials}
%\title[Non-equilibrium dynamics of kinematic tracers]{Non-equilibrium dynamics of kinematic tracers in time-dependent potentials}
%\title[Non-equilibrium dynamics of kinematic tracers]{Non-equilibrium dynamics of tracer particles in time-dependent potentials}
%\title[Non-equilibrium statistical mechanics]{Non-equilibrium mechanics of collisionless systems}
%\title[Non-equilibrium gravitational mechanics]{Non-equilibrium gravitational mechanics of collisionless systems}
%\title[Non-equilibrium gravitational mechanics]{Non-equilibrium gravitational mechanics}
\title[Non-equilibrium gravitational mechanics]{A probability theory for non-equilibrium gravitational systems}
\author[Jorge Pe\~{n}arrubia]{Jorge Pe\~{n}arrubia$^{1}$\thanks{jorpega@roe.ac.uk}\\
$^1$Institute for Astronomy, University of Edinburgh, Royal Observatory, Blackford Hill, Edinburgh EH9 3HJ, UK\\}
\maketitle  

\begin{abstract} 
This paper uses dynamical invariants to describe the evolution of collisionless systems subject to time-dependent gravitational forces without resorting to maximum-entropy probabilities. 
We show that collisionless relaxation can be viewed as a special type of diffusion process in the integral-of-motion space. In time-varying potentials with a fixed spatial symmetry the diffusion coefficients are closely related to virial quantities, such as the specific moment of inertia, the virial factor and the mean kinetic and potential energy of microcanonical particle ensembles. 
The non-equilibrium distribution function (DF) is found by convolving the initial DF with the Green function that solves Einstein's equation for freely diffusing particles. Such a convolution also yields a natural solution to the Fokker-Planck equations in the energy space. Our mathematical formalism can be generalized to potentials with a time-varying symmetry, where diffusion extends over multiple dimensions of the integral-of-motion space.
The new probability theory is in many ways analogous to stochastic calculus, with two significant differences: (i) the equations of motion that govern the trajectories of particles are fully deterministic, and (ii) the diffusion coefficients can be derived self-consistently from microcanonical phase-space averages without relying on ergodicity assumptions. 
For illustration we follow the cold collapse of $N$-body models in a time-dependent logarithmic potential. Comparison between the analytical and numerical results shows excellent agreement in regions where the potential evolution does not depart too strongly from the adiabatic regime.
\end{abstract}   

\begin{keywords}
Galaxy: kinematics and dynamics; galaxies: evolution; diffusion < Physical Data and Processes. 
\end{keywords}

\section{Introduction} \label{sec:intro}
Systems with a very large number of degrees of freedom exhibit regularities that can be expressed as statistical laws. For gases and plasmas it is possible to derive such laws in a reasonably straightforward manner from certain general principles of statistical mechanics (e.g. Landau \& Lifshitz 1980). Unfortunately, statistical mechanics has serious, unresolved issues when it comes to describing the non-equilibrium evolution of dynamical systems subject to long-range forces such as gravity (see Padmanabhan 1990 and Campa et al. 2009 for detailed reviews). 
Particles interacting via long-range forces exhibit some unusual behaviour, like non-ergodicity and negative specific heat (Antonov 1961; Lynden-Bell \& Lynden-Bell 1977; see Lynden-Bell 1999 for a review), which is difficult to incorporate into a classical thermodynamical framework. These difficulties are in part responsible for the current lack of theoretical understanding of how virialized systems settle in their particular equilibrium configuration.

The first attempt to construct a statistical theory for the evolution of collisionless galaxies is due to Lynden-Bell (1967), who used concepts of classical thermodynamics in order to predict the {\it final equilibrium} state of a collisionless gravitating system evolving in a time-dependent gravitational field, a process dubbed as `violent relaxation'. In his pioneer work Lynden-Bell discretizes phase space using elements (macrocells) of equal volume and different masses. Applying standard techniques of mechanical statistics he shows that violent relaxation leads toward a unique {\it coarse-grained} distribution function ($f_c$) which can be described by a weighted superposition of Fermi-Dirac distributions with different temperatures. Nakamura (2000) proposes an alternative approach where he derives the equilibrium distribution function of macrocells of equal mass and different phase-space volumes using Jaynes (1957) information theory. The new phase-space partition leads to a coarse-grained distribution function that follows the well-known Mawell-Boltzmann distribution. To date it remains unclear which of the two approaches is more correct from a statistical point of view (Arad \& Lynden-Bell 2005), although both fail to reproduce numerical experiments of violent relaxation (Arad \& Johansson 2005). 

%By the end of the $19^{\rm th}$ century the kinetic theory of gases revolutionized our physical undertanding of the composition of the world by stating that all macroscopic phenomena, such as temperature, volume, pressure, conductivity, viscosity,..., etc are manifestations of the mechanical interactions between microscopic entities, at the time known as `molecules'.  
%Unfortunately, by the end of the $20^{\rm th}$ century it became increasingly clear that statistical mechanics has serious, unresolved issues when it comes to describing the non-equilibrium behaviour of dynamical systems subject to long-range forces such as gravity (see Padmanabhan 1990 for a detailed review).
%the averaged properties of gravitating systems can be derived from general principles of statistical mechanics if these systems find themselves in a state of dynamical equilibrium. This explains why there still is a lack of theoretical understanding of how virialized systems settle in their particular equilibrium configuration.  
%In fact none of these theories provide a complete description of the equilibrium state of gravitating systems.
The central difficulty in accepting the distribution function derived from Lynden-Bell and Nakamura's theories is that it predicts infinite mass for the system. In other words, the variational problem that determines the most probable $f_c$ possesses no solution for any finite total mass. This shortcoming may be due to the short life of the process that drives relaxation, i.e. fluctuations of the gravitational field, which vanish on the time scale $(G \rho)^{-1/2}$, well before the thermodynamical equilibrium is attained. As a result, in most gravitating systems the evolution will be frozen in a subdomain of the available phase space (a.k.a. `incomplete relaxation'), breaking the ergodicity hypothesis, which establishes an equivalence between time average and average over particle ensembles. As a possible remedy, Robert \& Sommeria (1992) (see also Chavanis et al. 1996) propose the `maximum entropy production principle', which assumes that the evolution a system proceeds in such a way that the rate of entropy production is maximized while satisfying all the dynamical constraints. In this theory the coarse-grained distribution function is a solution to a modified version of the (collisionless) Fokker-Planck equation. These authors find that $f_c$ evolves toward a Fermi-Dirac distribution as the system approaches complete relaxation. More recently, there have been attempts to derive the most probable, maximum-entropy $f_c$ after imposing additional constraints on the system, e.g. adiabatic invariance of actions (Pontzen \& Governato 2013). 

Unfortunately, the concept of thermodynamical `entropy' ($\mathcal{H}$) is ill defined in systems subject to unshielded, long range forces. 
For ideal gases
 the $\mathcal{H}$-function must obey very special properties (see Appendix A of Jaynes 1957) which can only be met by the Boltzmann entropy, $\mathcal{H}=-\int f_c\ln f_c \d^3{\bf r}\d^3{\bf v}$. In contrast, for collisionless gravitating systems Tremaine et al. (1986) demonstrate that {\it any} convex function $C'(f_c)$ will lead to an entropy $\mathcal{H}=-\int C'\d^3{\bf r}\d^3{\bf v}$ that increases with time. In this regard the Boltzmann entropy $C'=f_c\ln f_c$ is only one of a wide variety of $\mathcal{H}$-functions that must increase monotonically during violent relaxation. 

This conclusion has been rejected by Dejonghe (1987) and Sridhar (1987) who provide counter-examples of systems in which $\mathcal{H}$ oscillates during mixing. Dejonghe (1987) concludes that Tremaine et al. theorem only proves that the $\mathcal{H}$-functions increase as a result of coarse-graining (which has an effect akin to information loss), and that 
whether or not the variation of $\mathcal{H}$ happens monotonically depends on (i) the prior information on the state of the system (e.g. the assignment of a priori probabilities to the macrocells) and (ii) the functional form of $C'(f_c)$. Since no particular choice can be justified from variational principles it remains unclear how to interpret thermodynamical entropies.
% More recently, Dehen (2005) puts the $\mathcal{H}$-theorem in a more solid footage by introducing an excess-mass function $D'(f)=\int (f_c-f)\d^3{\bf r}\d^3{\bf v}$ (here $f$ is the {\it fine-grained} distribution function), which always decreases on mixing in static potentials.

Padmanabhan (1990) identifies a different, but very worrisome limitation of thermodynamical approaches to violent relaxation. While in systems interacting via short-range forces the energy will be an {\it extensive} parameter (i.e. the total energy of the system will be the sum of the energies of the subsystems), in gravitating objects this cannot be guaranteed. Indeed, the extensive nature of the energy does not hold in the presence of long range interactions because 
the energy of a particle in a given macrocell cannot be shielded from the gravitational attraction of particles in neighbour cells. When this happens the system can no longer be divided into non-interacting subsystems and the laws of equilibrium thermodynamics do not apply. 
%Padmanabhan shows that for gravitationally-bound systems the use of canonical and microcanonical ensembles do lead to different physical predictions. 
Recent work by Levin et al. (2008, 2014) supports this conclusion. With the aid of $N$-body simulations these authors follow the evolution of self-gravitating models with `water-bag' distribution functions that are not initially in balance. The $N$-body models are left to oscillate with decreasing amplitude until they reach dynamical equilibrium. Levin et al. (see also Joyce et al. 2009 and Sylos Labini et al. 2015)
show that the final state after violent relaxation is typically characterized by a bi-modal distribution function, with particles in the inner-most region of the potential distributing according to Lynden-Bell's theory while those in the outer-most regions forming a distinct dilute halo which forms through resonances arising during the macroscopic oscillations of the system. 
%The number of particles in the halo depends strongly on the initial configuration of the system. If the initial distribution is constructed in such a way as to suppress macroscopic oscillations the resulting stationary state is found to be the one predicted by Lynden-Bell. 

%Clearly, the non-extensive nature of gravity introduces a number of unresolved problems when applying classical statistical mechanics to non-equilibrium gravitating systems, which are in part responsible for our lack of theoretical understanding of how virialized systems settle in their particular equilibrium configuration.  

This contribution seeks to build a probability theory for {\it non-equilibrium} gravitating systems 
which does not rely on thermodynamical arguments (e.g. phase-space discretization, maximum-entropy probabilities,...) or on ergodicity assumptions. Our goal is not only to describe the final state of violent relaxation but also all the intermediate stages (i.e. incomplete relaxation). 
We shall work in the collisionless, mean field limit, where the granularity in the system may be ignored, such that the (fine-grained) distribution function $f({\bf r}, {\bf v}, t)$ is assumed to be smooth over the mean interparticle distance of the system.

%conceptually simple. Section~\ref{sec:micro} starts by describing a coordinate transformation in which the explict time dependence of the gravitational field vanishes from the equations of motion. Pe\~narrubia (2013, hereinafter P13) shows that after carrying the inverse transformation the integrals of motion admitted by the static potential become dynamical invariants in the original coordinates. In \S\ref{sec:df} we follow similar arguments to describe the evolution of a system whose initial state is specified by the distribution function $f({\bf r},{\bf v},t_0)$. 

The method proposed below follows up the work of Pe\~narrubia (2013, hereinafter P13), who devises of a general coordinate transformation that removes the explicit time-dependence from the equations of motion of a collisionless system (see \S\ref{sec:micro}). By referring the phase-space locations of particles to a coordinate frame in which the potential remains `static' the dynamical effects introduced by the time evolution vanish.  
In \S\ref{sec:df} we apply similar arguments to derive an {\it invariant distribution function} from the initial state of the system at $t=t_0$. Transforming the invariant function back to the original integral-of-motion space yields the desired, non-equilibrium $f({\bf r},{\bf v},t_0+\tau')$, where $\tau'>0$ is a time interval of arbitrary length. Interestingly, we find that this transformation is analogous to a Green convolution of the invariant function with the solution to a diffusion equation, suggesting that collisionless relaxation can be viewed as a special type of diffusion process in the integral-of-motion space.
In \S\ref{sec:examples} we run a suite of $N$-body experiments which follows the evolution of tracer particles undergoing cold collapse in a time-dependent scale-free potential. We will show that the theory works well in regions where the potential evolution does not depart too strongly from the adiabatic regime.
The great virtue of scale-free potentials is that the diffusion coefficients can be expressed analytically in terms of virial quantities associated to microcanonical ensembles. Indeed, it will become evident that deriving the diffusion coefficients in arbitrary potentials is generally a very difficult task. In \S\ref{sec:dis} we discuss future prospects, like the analysis of self-gravitating systems with a time-varying symmetry and the possibility to incorporate the effects of random particle-particle encounters into our theory. For clarity Appendix A describes a number of concepts encountered in statistical mechanics, such as microcanonical ensembles, phase mixing, ergodicity,.., which appear repeatedly in the present work. Appendix B shows that Markov chains provide useful statistical tools to describe the evolution of collisionless systems.

\section{From micro to macro} \label{sec:micro}
Consider a microcanonical ensemble of particles that at the time $t=t_0$ move on orbits
with an energy $E=E_0$ in a time-dependent potential $\Phi({\bf r},t)$. In this Section we use dynamical invariants (i.e. quantities that are conserved along the phase-space path of a particle) to demonstrate that the
evolution of the microcanonical distribution can be described by the same phenomenological equations that arise in stochastic processes such as Brownian motion and random walks (see Wax 1958 for a review). 
%If we imagine a macroscopic system as the sum of an infinite number of microcanonical subsystems, then a derivation of the macrocanonical phase-space distribution naturally follows from the convolution of the initial distribution function with the microcanonical energy distribution.

\subsection{Dynamical invariants}\label{sec:inv}
The equations of motion of a particle subject to a time-dependent gravitational acceleration can be written as
\begin{equation}
\label{eq:eqmot}
\ddot{\bf r}= {\bf F}({\bf r},t),
\end{equation}
where ${\bf F}$ is a specific force.
If the force is conservative one can find  a gravitational potential $\Phi$ such that ${\bf F}({\bf r},t)=-{\bf \nabla}\Phi({\bf r},t)$.
%In what follows our calculations are derived in the mean-field limit, thus ignoring the granularity of $N$-body systems.

Lynden-Bell (1982) and P13 show that under these conditions one can construct a canonical transformation ${\bf r}\mapsto {\bf r}'R(t)$, and a time-coordinate transformation $\d t\mapsto \d \tau R^{2}(t)$, such that the explicit time-dependence of the force field vanishes from the equations of motion. In the new coordinates Equation~(\ref{eq:eqmot}) becomes
\begin{equation}
\label{eq:eqmot_tind}
\frac{\d^2 {\bf r}'}{d \tau^2}={\bf F}'[{\bf r}'].
\end{equation}

Equating~(\ref{eq:eqmot}) and~(\ref{eq:eqmot_tind}) shows that the scaling factor $R(t)$ must be a solution of the following differential equation
\begin{equation}
\label{eq:emar_R}
\ddot R R^3 {\bf r}'- R^3 {\bf F}(R{\bf r}',t)= -{\bf F}'[{\bf r}'].
\end{equation}
Note that if ${\bf F}$ is a conservative force (${\bf \nabla}\times{\bf F}=0$) then ${\bf F}'$ is also conservative. This calls for the definition of a time-independent scalar potential, $\Phi'=-\int F'\d r'$, such that in the transformed coordinates the energy, $I=1/2(\d {\bf r}'/\d \tau)^2+\Phi'({\bf r}')$, becomes an exact dynamical invariant (i.e. a constant of motion). Expressing the energy invariant $I$ in terms of the original coordinates yields
\begin{equation}
\label{eq:inv}
I = \frac{1}{2}(R {\bf v}- \dot R {\bf r})^2 + \frac{1}{2}\ddot R R r^2 + R^2\Phi({\bf r},t);
\end{equation}
where ${\bf v}=\d{\bf r}/\d t$. It is important to emphasize that the energy $I$ is conserved {\it along the phase-space path of a particle motion}, i.e. the subset of the $6-$dimensional phase space on which the equations of motion~(\ref{eq:eqmot}) are fulfilled.

Equation~(\ref{eq:inv}) can be easily re-arranged to express the energy as a function of the invariant quantity, that is
\begin{equation}
\label{eq:ei}
E=E_a+ ({\bf r}\cdot{\bf v})\frac{\dot R}{R}-\frac{1}{2}r^2 \bigg(\frac{\ddot R}{R}+\frac{\dot R^2}{R^2}\bigg);
\end{equation}
where $E_a(t)\equiv I/R^2(t)$ denotes the {\it adiabatic energy}. 
In what follows it is useful to define the quantity $\Delta({\bf r},{\bf v},t)\equiv E-E_a$ as the change of energy with respect to the adiabatic solution. In gravitational fields that do not vary rapidly with time the energy change is $|\Delta/E_a|\lesssim 1$, which defines the so-called adiabatic regime.
 Note that $\Delta$ can be positive or negative depending on the sign of ${\bf r}\cdot{\bf v}$, and that the tangential velocity component does not contribute to $\Delta$, which implies that the energy change is
independent of the angular momentum of the orbit.

\subsubsection{Scale-free potentials}\label{sec:scalefree}
Power-law gravitational fields are of special interest in this work, as they permit simple approximate solutions to Equation~(\ref{eq:emar_R}). Consider the time-dependent power-law force
\begin{equation}
\label{eq:plawf}
F(r,t)=-\mu(t)r^n.
\end{equation}
The potential associated with Equation~(\ref{eq:plawf}) can be written as
\begin{equation}
\label{eq:plawp}
\Phi(r,t)-\Phi_\infty= 
\begin{cases}
-\frac{\mu(t)}{n+1} r^{n+1} & ,n\ne - 1 \\ 
\mu(t)\ln(r)& , n=-1,
\end{cases}
\end{equation}
where $\Phi_\infty={\rm const.}$

Through a straightforward manipulation of Equation~(\ref{eq:emar_R}) P13 shows that the scale factor
\begin{equation}
\label{eq:R}
R(t)=\bigg[\frac{\mu(t)}{\mu_0}\bigg]^{-1/(n+3)},
\end{equation}
provides a solution of Equation~(\ref{eq:emar_R}) that is accurate at order $\mathcal{O}(\epsilon'^2)$, where $\mu_0=\mu(t_0)$, $\epsilon'=(\dot \mu/\mu_0) P$ and $P$ is the radial period of the orbit at $t=t_0$. From Equation~(\ref{eq:R}) it follows that $\dot R/R =-(\dot \mu/\mu)/(n+3)$. 
%Notice that $R$ and its time derivative are independent of the phase-space location where they are measured.

\subsection{The microcanonical ensemble}\label{sec:diff}
Let us now consider an ensemble of particles with energy $E=E_a$ at $t=t_0$. 
%For simplicity let the particles be isotropically distributed, so that the probability to find a particle in the phase-space volume $\d^6\Omega=\d^3{\bf r}\d^3{\bf v}$ centred at $({\bf r},{\bf v}$) at the time $t=t_0$ is given by 
%\begin{equation}
%\label{eq:R}
%f_E({\bf r},{\bf v},t_0)=\frac{\delta(E-H)\d^6\Omega}{\int \delta(E-H)\d^6\Omega}=\frac{1}{\omega(E,t_0)}\delta(E-H)\d^6\Omega,
%\end{equation}
%where $H=1/2v^2+\Phi({\bf r},t_0)$ is the Hamiltonian of the system and $\omega$ is the so-called density of states (see Appendix~\ref{sec:aver}).
A derivation of the energy distribution at an arbitrary time $t=t_0+\tau'$ may appear as an impossibly difficult task given that the value of $\Delta=E-E_a$ varies according to the phase-space location of each individual particle in the ensemble. To attack this problem we shall follow a statistical method originally introduced by Einstein (1905)\footnote{An English translation can be found in Stachel (1998).}, which is commonly applied to a wide range of stochastic processes encountered in Physics, Biology and Chemistry. This approach requires that the evolution of the gravitational field happens in the linear regime and does not deviate too strongly from the adiabatic limit.

As a first step one defines $\varphi(\Delta|E_a)\d \Delta$ as the (conditional) probability that during a time interval $(t_0, t_0+\tau')$ a particle with energy $E_a$ will experience an energy change $\Delta$ within the range $\d \Delta$. The function $\varphi$ is known as the {\it probability of energy change} and it is normalized such that
\begin{equation}
\label{eq:norm}
\int\varphi(\Delta|E_a)\d \Delta=1.
\end{equation} 
Since $\Delta=\Delta({\bf r},{\bf v},t)$ the probability function $\varphi$ must be intimately related to the distribution of particles in phase space, an issue that is discussed in detail in Section~\ref{sec:coeff}. 

As a second step one introduces the function $p(E,t|E_a,t_0)$, which defines the probability that a particle with energy $E_a$ at $t=t_0$ will have an energy in the range $E,E+\d E$ at the time $t=t_0+\tau'$. Since the number of particles is conserved these two functions must obey the following equality
\begin{eqnarray}
\label{eq:pe0}
%p(E,t_0+ \tau')%\d^3{\bf r}\d^3{\bf v}
%=\d^3{\bf r}\d^3{\bf v}\int p(E-\Delta,t_0)\varphi(\Delta)\d\Delta.
p(E,t_0+ \tau'|E_a,t_0)\d E
=\d E \int p(E-\Delta,t_0|E_a,t_0)\varphi(\Delta|E_a)\d\Delta,
\end{eqnarray} 
which is typically known as {\it Einstein's master equation}. 

If the evolution of the potential does not deviate too strongly from the adiabatic limit, i.e. $|\Delta/E_a|\lesssim 1$, one can expand the right-hand side of Equation~(\ref{eq:pe0}) in series of $\Delta$, which after taking into account the normalization~(\ref{eq:norm}) yields
\begin{eqnarray}
\label{eq:pe1}
p(E,t_0+ \tau'|E_a,t_0)\simeq p(E,t_0|E_a,t_0) 
- \frac{\partial p}{\partial E}\bigg|_{\Delta=0}\int\varphi(\Delta|E_a)\Delta \d \Delta \\ \nonumber
+\frac{1}{2}\frac{\partial^2 p}{\partial E^2}\bigg|_{\Delta=0}\int\varphi(\Delta|E_a)\Delta^2\d \Delta.
\end{eqnarray} 
Given that the time interval can be arbitrary small, the left-hand side of Equation~(\ref{eq:pe0}) can be also approximated as
\begin{equation}
\label{eq:pe2}
p(E,t_0+\tau'|E_a,t_0)\simeq p(E,t_0|E_a,t_0) + \frac{\partial p}{\partial t}\tau',
\end{equation} 
Equating (\ref{eq:pe1}) and~(\ref{eq:pe2}) yields 
\begin{eqnarray}
\label{eq:pe3}
\frac{\partial p}{\partial t}=C\frac{\partial p}{\partial E}\bigg|_{\Delta=0}+D\frac{\partial^2 p}{\partial E^2}\bigg|_{\Delta=0};
\end{eqnarray} 
where the {\it drift} coefficient is defined as
\begin{equation}
\label{eq:C}
C(E_a,t)=-\frac{1}{\tau'}\int\varphi(\Delta|E_a)\Delta \d \Delta,
\end{equation}
and the {\it diffusion} coefficient as
\begin{equation}
\label{eq:D}
D(E_a,t)=\frac{1}{2\tau'}\int\varphi(\Delta|E_a)\Delta^2 \d \Delta.
\end{equation}
Introducing the time-dependent variable $E'=E+C t$ reduces Equation~(\ref{eq:pe3}) to Einstein's well-known diffusion equation
\begin{eqnarray}
\label{eq:diff}
\frac{\partial p}{\partial t}=D\frac{\partial^2 p}{\partial E'^2}\bigg|_{\Delta=0}.
\end{eqnarray} 

By construction, the initial conditions of the ensemble demand that $p(E,t\to t_0|E_a,t_0)=\delta(E-E_a)$. This problem coincides with Einstein's diffusion outwards from a point, where $E-E_a$ plays the role of a `displacement' from the initial energy $E=E_a$. The general solution to Equation~(\ref{eq:diff}) is a Green function (e.g. Krapivsky et al. 2010)
\begin{equation}
\label{eq:p}
p(E,t|E_a,t_0)=\frac{1}{\sqrt{4\pi\tilde D(E_a,t)}}\exp\bigg\{-\frac{ [E-E_a+ \tilde C(E_a,t)]^2}{4 \tilde D(E_a,t)} \bigg\};
\end{equation} 
with $\tilde C=C\tau'$ and $\tilde D=D\tau'$. The mean and the dispersion of the distribution are $\overline{\Delta}=\overline{E-E_a}=-\tilde C$, and $\overline{\Delta^2}-\overline{\Delta}^2=2\tilde D$, respectively.
If the potential evolves adiabatically both coefficients $\tilde C$ and $\tilde D$ approach zero, and the probability function~(\ref{eq:p}) becomes sharply peaked about $E=E_a$. In the limit $\dot R=\ddot R\to 0$, one has $p(E,t|E_a,t_0)\to \delta (E-E_a)$, which recovers the adiabatic solution exactly. Section~\ref{sec:coeff} discusses the physical meaning of these coefficients and explores cases where $\tilde C$ and $\tilde D$ can be derived analytically. 

The evolution of a microcanonical ensemble in a time-dependent gravitational potential is therefore governed by Einstein's equation for freely diffusing particles. A few aspects of this result are worth highlighting. First, notice that the distribution~(\ref{eq:p}) is centred at $\overline\Delta=\overline{E-E_a}=-\tilde C$. We shall see below that $\tilde C\ne 0$ for particle ensembles that have not in reached a state of dynamical equilibrium. 
Second, in contrast to the Brownian motion here the energy change $\Delta$ does not arise from a stochastic mechanical exchange of energy with a granular medium but from the smooth time dependence of the gravitational field.
%Eq.~(\ref{eq:p}), therefore, can be regarded as a statistical description of a continuos process. 
This difference, albeit subtle, has very important consequences. Whereas in Einstein's original paper one finds that the mean squared displacement of Brownian particles grows as $\overline {x^2}=2 D \tau'\propto (T/k)\tau'$, where $T$ corresponds to the temperature of the system and $k$ to the viscosity coefficient, here the mean squared energy change with respect to the adiabatic solution, $\overline{\Delta^2}$, does not depend explicitly on $\tau'$. 
Indeed, the fact that $\tau'$ vanishes from Equation~(\ref{eq:p}) can be traced back to the {\it continuous} time-dependence of the gravitational field, which removes the problem encountered by Einstein to describe the discontinuous nature of the Brownian phenomenon during very short time intervals\footnote{Indeed, in Einstein (1905) the coefficient $D\propto \langle x^2\rangle/\tau'$ diverges in the limit $\tau'\to 0$. It became clear later on that during very short time intervals the Brownian motion does not behave like a Markovian process.}. 
%The fact that $\tau'$ vanishes from Equation~(\ref{eq:p}) can be traced back to the assumption that a change in the gravitational field leads to an {\it instantaneous} change in the orbital energy of the particles. Such an approximation, which is accurate in a non-relativistic regime, removes the problem encountered by Einstein to describe the discontinuous nature of the Brownian phenomenon during very short time intervals\footnote{Indeed, the coefficient $D\propto \langle x^2\rangle/\tau'$ diverges in the limit $\tau'\to 0$. It became clear later on that the Brownian motion cannot be assumed to behave like a Markovian process in this limit.}. 
%The distinction between stochastic and deterministic processes needs to be emphasized, as the above results indicate that the evolution in phase-space of systems subject to a time-dependent gravitational force will be different depending on whether the change of energy of individual particles is driven by a continuous dynamical process, or by a random stochastic one, even though the equations that describe the evolution share the same form, i.e. that of Einstein's diffusion equation.
The distinction between stochastic and deterministic processes 
needs to be emphasized, as the above results indicate that the evolution in phase space of systems subject to long-range forces will depend on whether the energy change is driven by a continuous dynamical process or by a random stochastic one, even though the equations that describe the evolution share the same form, i.e. that of Einstein's diffusion equation.
%Therefore, whether the change of energy of individual particles is driven by a continuous dynamical process or by a random stochastic one leads to a different physical behaviour of the system, even though the equations that describe its dynamical evolution share the same form, i.e. that of Einstein's diffusion equation.

\begin{table*}
\centering
\begin{minipage}{160mm}
 \begin{tabular}{lccc}
 \hline
 & $n<-1$ & $n=-1$ & $n>-1$  \\\hline
Phase-space volume p.u. energy ($w$)& $\frac{(4 \pi)^2\sqrt{2\pi}}{3}\frac{-(n+1)}{7+n}\frac{\Gamma[1/2-3/(n+1)]}{\Gamma[-3/(n+1)]}|E|^{1/2}r_m^3$  & $(4\pi)^2\sqrt{\frac{\pi}{54}}\mu^{1/2}r_m^3$ & $\frac{(4\pi)^2\sqrt{2\pi}}{6}\frac{\Gamma[1+3/(n+1)]}{\Gamma[3/2+3/(n+1)]}|E|^{1/2}r_m^3$ \\
Constant ($B_n$)& $\frac{2}{3}\frac{(n+1)(7+n)}{7-3 n}\frac{\Gamma[1/2-3/(n+1)]}{\Gamma[3/2-3/(n+1)]}\frac{\Gamma[2-3/(n+1)]}{\Gamma[-3/(n+1)]}$ & $\frac{6}{5}$ & $\frac{2\Gamma[5/2+3/(n+1)]}{\Gamma[3/2+3/(n+1)]}\frac{\Gamma[3/2+5/(n+1)]}{\Gamma[5/2+5/(n+1)]}$ \\ 
Specific moment of inertia ($\mathcal{I}$) & $\frac{3}{10}\frac{7+n}{11+n}\frac{\Gamma[1/2-5/(n+1)]}{\Gamma[1/2-3/(n+1)]}\frac{\Gamma[-3/(n+1)]}{\Gamma[-5/(n+1)]}r_m^2$ & $\frac{1}{2}\big(\frac{3}{5}\big)^{3/2}r_m^2$ & $\frac{1}{2}\frac{\Gamma[1+5/(n+1)]}{\Gamma[1+3/(n+1)]}\frac{\Gamma[3/2+3/(n+1)]}{\Gamma[3/2+3/(n+1)]}r_m^2$ \\
Mean kinetic energy ($T$) & $\frac{3}{5-n}\frac{\Gamma[3/2-3/(n+1)]}{\Gamma[1/2-3/(n+1)]}\frac{\Gamma[-3/(n+1)]}{\Gamma[2-3/(n+1)]}|E|$ & $\frac{\mu}{6}$ & $\frac{1}{2}\frac{\Gamma[3/2+3/(n+1)]}{\Gamma[5/2+3/(n+1)]}|E|$ \\
\hline
\end{tabular}
\end{minipage}
\caption[]{Quantities defining the isotropic coefficient $\tilde D(E,t)$ in Equation~(\ref{eq:D3}) derived in a gravitational field $F(r,t)=\mu(t) r^n$. Here $E$ is the energy of the microcanonical ensemble, and $r_m(E,t)$ is the maximum radius that the particles with this energy can reach, that is $\Phi(r_m,t)=E$, with $\Phi$
%=\mu(t) r^{n+1}/(n+1)$ 
given by Equation~(\ref{eq:plawp}). The Gamma function is defined as $\Gamma(z)=\int_0^\infty t^{z-1}\exp[-t]\d t$.}
\end{table*}
%with the specific moment of inertia
%\begin{equation}
%\label{eq:Iplaw}
%\mathcal{I}= 
%\begin{cases}
%\frac{1}{2}\frac{\Gamma[1+5/(n+1)]}{\Gamma[1+3/(n+1)]}\frac{\Gamma[3/2+3/(n+1)]}%{\Gamma[3/2+3/(n+1)]}r_m^2\ & ,n>-1 \\ 
%\frac{3}{10}\frac{7+n}{11+n}\frac{\Gamma[1/2-5/(n+1)]}{\Gamma[1/2-3/(n+1)]}\frac%{\Gamma[-3/(n+1)]}{\Gamma[-5/(n+1)]}r_m^2\ & ,n<-1 \\ 
%\frac{1}{2}\big(\frac{3}{5}\big)^{3/2}r_m^2& , n=-1,
%\end{cases}
%\end{equation}
%the kinetic energy per degree of freedom
%\begin{equation}
%\label{eq:Tplaw}
%T=  
%\begin{cases}
%\frac{1}{2}\frac{\Gamma[3/2+3/(n+1)]}{\Gamma[5/2+3/(n+1)]}E & ,n>-1 \\ 
%\frac{3}{5-n}\frac{\Gamma[3/2-3/(n+1)]}{\Gamma[1/2-3/(n+1)]}\frac{\Gamma[-3/(n+1%)]}{\Gamma[2-3/(n+1)]}E\ & ,n<-1 \\ 
%\frac{\mu}{6}\ & , n=-1,
%\end{cases}
%\end{equation}
%and $B_n$ is a positive constant
%that only depend on the power-law index of the force field~(\ref{eq:plawf}) as
%\begin{equation}
%\label{eq:Bplaw}
%B_n=  
%\begin{cases}
% 2\frac{\Gamma[5/2+3/(n+1)]}{\Gamma[3/2+3/(n+1)]}\frac{\Gamma[3/2+5/(n+1)]}{\Gam%ma[5/2+5/(n+1)]}& ,n>-1 \\ 
%\frac{2}{3}\frac{(n+1)(7+n)}{7-3 n}\frac{\Gamma[1/2-3/(n+1)]}{\Gamma[3/2-3/(n+1)%]}\frac{\Gamma[2-3/(n+1)]}{\Gamma[-3/(n+1)]}\ & ,n<-1 \\ 
%\frac{6}{5}\ & , n=-1,
%\end{cases}
%\end{equation}
%where $\Gamma(z)=\int_0^\infty t^{z-1}\exp[-t]\d t$ is the Gamma function.

\subsection{Drift and diffusion coefficients }\label{sec:coeff}
Let us now turn to the thorny task of calculating the coefficients appearing in Equation~(\ref{eq:p}). An analytical derivation of $\tilde C$ and $\tilde D$ is generally not possible due to the coupling between Equation~(\ref{eq:emar_R}) and the equations of motion~(\ref{eq:eqmot}), which introduces a dependence in the scale factor on the initial conditions of the orbit.
%, that is $R=R(t,E_0,\{\alpha\}_{\nu,0})$.
However, for isotropic particle ensembles orbiting in scale-free gravitational fields it is possible to derive approximate solutions that offer useful insight into the physical meaning of these coefficients. In particular, it will be shown below that both $\tilde C$ and $\tilde D$ are closely related to quantities appearing in the virial theorem.

We start by recalling that the probability function $\varphi(\Delta,E_a)$ is proportional to the number of particles in phase-space experiencing an energy change $E-E_a=\Delta({\bf r},{\bf v},t)$. Hence, Equations~(\ref{eq:C}) and~(\ref{eq:D}) can be written, respectively, as statistical averages of $\Delta$ and $\Delta^2$ over the phase-space volume $\omega(E,t)$ accessible to particles with energy $E$, as discussed in Appendix~\ref{sec:aver}. Using Equation~(\ref{eq:xam}) the coefficients $\tilde C$ and $\tilde D$ can be calculated as
\begin{equation}
\label{eq:C2}
\tilde C(E,t)\approx -\frac{1}{\omega}\int (\dot R/R)({\bf r}\cdot{\bf v}) \delta(E-H)\d^3{\bf r}\d^3{\bf v},
\end{equation}
and
\begin{equation}
\label{eq:D2}
\tilde D(E,t)\approx \frac{1}{2\omega}\int (\dot R/R)^2({\bf r}\cdot{\bf v})^2\delta(E-H)\d^3{\bf r}\d^3{\bf v},
\end{equation}
where we have neglected all terms at order $\ddot R$ and $\dot R^2$ in Equation~(\ref{eq:ei}), so that $\Delta=E-E_a\approx (\dot R/R)({\bf r}\cdot{\bf v})$. This approximation is fairly accurate in potentials that do not evolve rapidly with time (see the numerical tests of P13).

\subsubsection{Scale free potentials}\label{sec:coeffplaw}
In scale-free gravitational fields the scale factor $R$ is independent of the phase-space trajectory of individual particles and only depends on the time $t$, hence Equations~(\ref{eq:C2}) and~(\ref{eq:D2}) can be simply written as $\tilde C\simeq -(\dot R/R) \langle {\bf r}\cdot{\bf v} \rangle $, and $\tilde D\simeq (\dot R/R)^2\langle ({\bf r}\cdot{\bf v})^2 \rangle $, with the brackets denoting microcanonical phase-space averages (see Appendix~\ref{sec:aver}). 
Note that $\tilde C$ is proportional to $ \langle {\bf r}\cdot{\bf v} \rangle $, a quantity that is typically known as the {\it virial function}. This calls for the definition of a specific moment of inertia
\begin{eqnarray}
\label{eq:Imom}
\mathcal{I}(E,t)&=&\frac{1}{2\omega}\int r^2 \delta(E-H) \d^3{\bf r}\d^3{\bf v}\\ \nonumber
&=&\frac{(4\pi)^2}{2 \omega}\int_0^{r_m} r^4[2(E-\Phi)]^{1/2}\d r,
\end{eqnarray}
and a mean kinetic energy per degree of freedom
\begin{eqnarray}
\label{eq:T}
T(E,t)&=&\frac{1}{\omega}\int \frac{{\bf v}^2}{2} \delta(E-H) \d^3{\bf r}\d^3{\bf v}\\ \nonumber
&=&\frac{(4\pi)^2}{6 \omega}\int_0^{r_m} r^2[2(E-\Phi)]^{3/2}\d r,
\end{eqnarray}
where $r_m(E,t)$ is the maximum radius that particles with energy $E$ can reach, that is $\Phi(r_m,t)=E$.

From the definition of the specific moment of inertia it is straightforward to show that $\mathcal{ I}=\langle {\bf r}^2/2\rangle$, $\mathcal{ \dot I}=\langle {\bf r}\cdot{\bf v} \rangle$ and $\mathcal{\ddot I}=\langle v^2 \rangle + \langle {\bf r}\cdot{\bf F} \rangle= 2 T + W$, where $W(E,t)$ is the averaged potential energy of the microcanonical distribution (e.g. Heggie \& Hut 2003). A state of {\it dynamical equilibrium} is therefore attained when distribution in configuration space obeys the virial theorem, $\mathcal{\dot I}=\mathcal{\ddot I}= 2T+W=0$. 

The drift coefficient~(\ref{eq:C2}) can be written in terms of virial quantities as
\begin{equation}
\label{eq:C3}
\tilde C(E,t) = -(\dot R/R)\mathcal{\dot I},
\end{equation}
which shows that the diffusion process outlined in \S\ref{sec:diff} is closely related to the virializiation of systems out of equilibrium. 
Indeed, taking the time derivative on both sides of Equation~(\ref{eq:C3}) and neglecting terms at order $\ddot R$ and $\dot R^2$ yields
\begin{equation}
\label{eq:Ct}
\frac{\partial \tilde C}{\partial t}\approx -(\dot R/R)\ddot I=-(\dot R/R)(2T+W).
\end{equation}
Microcanonical ensembles in virial equilibrium at $t=t_0$ obey $\mathcal {\dot I}\simeq \mathcal{\ddot I}\simeq 0$, which is equivalent to the condition $\tilde C\simeq \partial \tilde C/\partial t\simeq 0$. Taylor expanding the coefficient $\tilde C$ at the time $t=t_0+\tau'$ therefore yields $\tilde C(t)\approx \tilde C(t_0)+(\partial \tilde C/\partial t)\tau'\approx 0$, which substituted in Equation~(\ref{eq:p}) leads to $\overline{E}(t)\approx E_a(t)$. 
%We find therefore that the mean energy of a particle ensemble that is initially in virial equilibrium will evolve in the adiabatic limit. 
This is an important result, as we find that systems that are initially virialized follow an evolution that is close to the adiabatic limit at all times. We shall say that these systems evolve in {\it quasi-dynamical equilibrium}.

Unfortunately, the derivation of $\tilde C$ in systems that are neither phase mixed\footnote{In this paper the term `mixed' is used in a fairly broad context. In particular, we shall say that particle ensembles are {\it unmixed} if they do not sample the accessible phase-space volume. An equivalent statement is that mixing is complete when phase-space occupation is maximum. Note that (i) the process of mixing happens in static as well as in time-dependent potentials, (ii) unmixed ensembles in general do not obey the ergodic principle.} nor virialized at $t=t_0$ requires precise information on the trajectories of individual particles in phase space, which in general renders the problem analytically intractable. Tidal streams provide a particularly handy example of systems that are not in dynamical equilibrium and do not mix efficiently (see discussion in P13).
%Systems out of dynamical equilibrium evolving in time-dependent potentials exhibit oscillations in the value of $\tilde C$ which lead to periodic fluctuations of the mean energy about the adiabatic value (see the $N$-body experiments of P13 for an illustration of this process).
Let us further illustrate the relation between phase mixing and virialization by discussing different states of a system at $t=t_0$ (we shall come back to these examples in \S\ref{sec:examples}). First, one can distribute particles in phase space such that the number of them moving away from the centre of the potential equals the number moving toward it. By construction, this ensemble is in dynamical balance, i.e. $\langle {\bf r}\cdot{\bf v} \rangle=\mathcal{\dot I}=0$ at $t=t_0$. Whether or not the system is also in virial equilibrium depends on the ratio between the kinetic and potential energies.
As discussed above, if the system is initially virialized ($\mathcal{\ddot I}= 2T+W=0$) the evolution occurs in the limit of quasi-dynamical equilibrium, so that $\tilde C\approx 0$ at $t=t_0+\tau'$.
However, if the condition for virial equilibrium is not satisfied (i.e. $\mathcal{\ddot I}\ne 0$) the system will evolve away from its initial configuration towards an equilibrium state. During the process the drift coefficient $\tilde C$ will exhibit periodic fluctuations which will damp progressively as the system regains its dynamical balance. 
A very similar process will be observed in systems whose initial configuration is not phase mixed (for example in tidal streams, where all particles are released from a progenitor galaxy with a wide range of velocities). In both cases collisionless relaxation proceeds in such a way that $\lim_{t\to \infty}\tilde C=0$. 
%Indeed, both virialization and phase-space mixing enter in Equation~(\ref{eq:C3}) through the diffusion process outlined in \S\ref{sec:diff}.

In particle ensembles that are phase mixed the diffusion coefficient $\tilde D$ can be analytically expressed as function of virial quantities
\begin{equation}
\label{eq:D3}
\tilde D(E,t)=B_n(\dot R/R)^2\mathcal{ I}T,
\end{equation}
where $B_n$ is a positive constant that only depends on the power-law index of the force~(\ref{eq:plawf}). Table~1 provides analytical expressions for $B_n$, $\mathcal{ I}$ and $T$ for different power-law indices. 
The coefficient $\tilde D$ is always positive and has a well-defined radial dependence which can be straightforwardly derived by inserting the values of Table~1 into Equation~(\ref{eq:D3}). The kinematic energy scales as $T\propto E=\Phi(r_m)\propto r_m^{n+1}$ and the specific moment of inertia as $\mathcal{I}\propto r_m^2$, hence the diffusion coefficient has a well-defined radial dependence $\tilde D\propto r_m^{3+n}$. For indices $n>-3$, which covers the range of potentials of astronomical interest, $\tilde D$ becomes very small at the centre of the potential and increases monotonically away from it, which implies that orbital diffusion predominantly affects the outer regions of gravitating systems.

\section{Time-evolution of the Distribution Function}\label{sec:df}
%Section~\ref{sec:diff} provides the necessary tools to map an initial state of the system to a corresponding future state at a later time.
Section~\ref{sec:diff} relies on the construction of energy invariants to derive the evolution of a microcanonical ensemble of particles. Here we use those results to describe the non-equilibrium state of a system composed of an infinite number of microcanonical subsystems.
 
Consider a system whose initial state is defined by the distribution function $f(E_0,\{\alpha\}_\nu,t_0)$, where $\{\alpha\}_\nu=\alpha_1,...,\alpha_\nu$ are $\nu$-integrals of motion related to the symmetry of the potential. 
If the normalization is chosen such that $\int f\d^3{\bf r}\d^3{\bf v}=1$, then $f(E_0[{\bf r},{\bf v}],\alpha_1[{\bf r},{\bf v}],...,\alpha_\nu[{\bf r},{\bf v}],t_0)\d^3{\bf r}\d^3{\bf v}$ defines  the probability to find a particle in the phase-space volume $\d^3{\bf r}\d^3{\bf v}$ centred at the coordinates ${(\bf r},{\bf v})$ at the time $t_0$. 
%By definition, particle ensembles orbiting in a time-varying potential cannot be found in a steady state, hence $\partial f/\partial t \ne 0$. 
In what follows it is convenient to work in the space of the integrals of motion. To this end we introduce the probability function 
$$N(E,\{\alpha\}_\nu,t)=f(E,\{\alpha\}_\nu,t)\omega(E,\{\alpha\}_\nu,t),$$
where $\omega$ is the density of states (see Appendix~\ref{sec:aver} for details). 

Our goal is to derive $N(E,\{\alpha\}_\nu,t)$ at $t=t_0+\tau'$ from the initial distribution $N(E_0,\{\alpha\}_\nu,t_0)$ in a time-dependent potential $\Phi({\bf r},t)$. This shall be done in two steps. In Section~\ref{sec:invdf} we derive an invariant probability distribution from the initial distribution function. In Section~\ref{sec:timedf} we use the invariant distribution in order to find the distribution of particles in the original integral-of-motion space at an arbitrary time $t=t_0+\tau'$.

%Clearly, $\d N/\d t=(\partial N/\partial I)\dot I + \sum_{i=1}^{\nu}(\partial N/\partial \alpha_\nu)\dot \alpha_i=0$, which implies that in this space the distribution function is invariant under the transformation $t=t_0+\tau'$. Thus $N(I,\alpha_\nu)$ does not contain an explicit time dependence. 

\subsection{Invariant Distribution}\label{sec:invdf}
In Section~\ref{sec:inv} we highlighted the fact that the canonical transformation that removes the explicit time-dependence from the equations of motion leaves the angular momentum invariant. This result implies that in time-dependent potentials which preserve the initial spatial symmetry the space defined by the constants of motion has $\nu+1$ dimensions, namely the invariant energy $I$ plus the $\nu-$integrals $\alpha_i$. The Jeans theorem states that the probability function expressed as a function of the constants of the motion $\{I,\alpha_1,...,\alpha_\nu\}$ yields a steady-state solution of the collisionless Boltzmann equation, i.e. $\d N/\d t=(\partial N/\partial I)\dot I + \sum_{i=1}^{\nu}(\partial N/\partial \alpha_\nu)\dot \alpha_i=\partial N/\partial t=0$. An equivalent statement is that the probability function $N(I,\alpha_1,...,\alpha_\nu,t_0)$ remains invariant under the transformation $t_0\mapsto t=t_0+\tau'$. In what follows we emphasize the latter property by removing the explicit time-dependence from the invariant distribution, which shall be denoted as $N(I,\{\alpha\}_\nu)$. In this work we are particularly interested in non-rotating spherical systems, which sets the integrals $(\alpha_1,\alpha_2,\alpha_3)=(L_x,L_y,L_z)$, noting in passing that in potentials with time-varying spatial symmetry the diffusion process outlined in \S\ref{sec:diff} takes place in multiple dimensions (see discussion in \S\ref{sec:dis}). 

Let us now derive the invariant distribution $N(I,\{\alpha\}_\nu)$ from the initial probability function $N(E_0,\{\alpha\}_\nu,t_0)$. Using Equation~(\ref{eq:ei}) the invariant energy can be written as $I=R_0^2(E_0-\Delta)$, where $R_0\equiv R(t_0)$. 
%We are still free to choose the initial value of the scaling function, $R(t_0)$. The simplest choice corresponds to $R(t_0)=1$, which sets the energy invariant $I=E_0-\Delta$. 
Following the same steps as in Section~\ref{sec:micro} it is straightforward to show that the probability that a particle with an energy $E_0$ at the time $t=t_0$ has an energy in the interval $I,I+\d I$ is
\begin{equation}
\label{eq:pi}
p(I|E_0,t_0)=\frac{1}{\sqrt{4\pi\tilde D(E_0,t_0)R_0^4}}\exp\bigg\{-\frac{ [I-E_0R_0^2- \tilde C(E_0,t_0)R_0^2]^2}{4 \tilde D(E_0,t_0)R_0^4} \bigg\};
\end{equation} 
with the coefficients $\tilde C$ and $\tilde D$ given by Equations~(\ref{eq:C}) and~(\ref{eq:D}), respectively. 

The probability function $p(I|E_0,t_0)$ is a solution to the differential Equation~(\ref{eq:diff}) and therefore a Green function (e.g. Krapivsky et al. 2010). Hence, the invariant probability function is found by convolving the initial distribution function $N(E_0,\{\alpha\}_\nu,t_0)$ with the probability function~(\ref{eq:pi}) 
\begin{eqnarray}
\label{eq:N1}
N(I,\{\alpha\}_\nu)= \int p(I|E_0,t_0)N(E_0,\{\alpha\}_\nu,t_0)\d E_0 \\ \nonumber
=\large\int \frac{N(E_0,\{\alpha\}_\nu,t_0)}{\sqrt{4 \pi \tilde D(E_0,t_0)R_0^4}}\exp\bigg\{-\frac{[I-E_0R_0^2-\tilde C(E_0,t_0)R_0^2]^2}{4 \tilde D(E_0,t_0)R_0^4}\bigg\}\d E_0,
\end{eqnarray} 
with the limits of the integral defined by the range of energies of the system (see examples in \S\ref{sec:examples}).
%Note that the above convolution conserves the total probability, i.e. $\int N(E,\{\alpha\}_\nu,t)\d E= \int N(E',\{\alpha\}_\nu,t_0)\d E'\int p(E,t|E',t_0)\d E=\int N(E',\{\alpha\}_\nu,t_0)\d E'=1$.

\subsection{Time-dependent Distribution Function}\label{sec:timedf}
One can now derive $N(E,\{\alpha\}_\nu,t)$ from the initial probability function $N(E_0,\{\alpha\}_\nu,t_0)$ following similar steps as 
as in \S\ref{sec:invdf}. 

Recall that the energy can be written as $E=E_a(t)+\Delta$, where $E_a\equiv I/R^2$ is the adiabatic energy. Let us introduce the function $N_a(E_a,\{\alpha\}_\nu,t)$ which defines the probability to find a particle within the adiabatic energy interval $E_a, E_a+\d E_a$ at the time $t$. In potentials with a fixed symmetry the probability function at the time $t=t_0+\tau'$ will be the result of the following Green convolution 
\begin{eqnarray}
\label{eq:Nt1}
N(E,\{\alpha\}_\nu,t)= \int p(E,t|E_a,t_0)N_a(E_a,\{\alpha\}_\nu,t_0)\d E_a,
\end{eqnarray} 
where the Green function $p$ is given by Equation~(\ref{eq:p}). The derivation of $N_a$ requires one extra Green convolution
\begin{eqnarray}
\label{eq:Nta_scale}
N_a(E_a,\{\alpha\}_\nu,t)= \int p(E_a,t|I)N(I,\{\alpha\}_\nu)\d I.
\end{eqnarray} 
In scale-free fields the probability distribution 
$p(E_a,t|I)=\delta(E_a-I/R^2)$ because $R$ does not have an intrinsic orbital dependence (see \S\ref{sec:scalefree}). In this case $N_a$ can be directly obtained from the invariant function~(\ref{eq:N1}) as
\begin{eqnarray}
\label{eq:Nta}
 N_a(E_a,\{\alpha\}_\nu,t)\d E_a=N(I[E_a],\{\alpha\}_\nu)\bigg|\frac{\d E_a}{\d I}\bigg|R^2 \d E_a=N(I,\{\alpha\}_\nu)\d I.
\end{eqnarray} 
Substituting~(\ref{eq:Nta}) into~(\ref{eq:Nt1}), inserting Equation~(\ref{eq:N1}) and re-arranging the integrands yields
\begin{eqnarray}
\label{eq:Nt2}
N(E,\{\alpha\}_\nu,t)=\int p(E,t|I R^{-2},t_0)N(I,\{\alpha\}_\nu)\d I\\ \nonumber
=\int \d E_0 N(E_0,\{\alpha\}_\nu,t_0)\int p(E,t|I R^{-2},t_0) p(I|E_0,t_0)\d I \\ \nonumber
\equiv \int \d E_0 N(E_0,\{\alpha\}_\nu,t_0)p_c(E,t|E_0,t_0).
\end{eqnarray}  
The probability function $p_c(E,t|E_0,t_0)$ is the result of the convolution of two Gaussians. In general, this function cannot be expressed analytically owing to the non-trivial dependence of the function $p(E,t|I R^{-2},t_0)$ on the energy $I$. However, in potentials that evolve slowly the function $p_c$ will peak sharply around $E\approx E_a=I/R^{2}$. Replacing the coefficients $\tilde C(I R^{-2},t)$ and $\tilde D(I R^{-2},t)$ on the right-hand integral of Equation~(\ref{eq:Nt2}) by $\tilde C(E,t)$ and $\tilde D(E,t)$, respectively, yields an analytical expression for $p_c$ which, not surprisingly, also is a Gaussian function
\begin{eqnarray}
\label{eq:pc}
p_c(E,t|E_0,t_0)&=&\int p(E,t|I R^{-2},t_0) p(I|E_0,t_0)\d I \\ \nonumber
&\approx& \frac{1}{\sqrt{4\pi\tilde D_c}}\exp\bigg\{ -\frac{[E - E_0 (R_0/R)^{2} +\Delta \tilde C]^2}{4\tilde D_c}\bigg\},
\end{eqnarray} 
where
\begin{eqnarray}
\label{eq:cc}
\Delta \tilde C(E,t,E_0,t_0)\equiv \tilde C(E,t)-\tilde C(E_0,t_0)\bigg(\frac{R_0}{R}\bigg)^2,
\end{eqnarray}
 and
\begin{eqnarray}
\label{eq:dc}
\tilde D_c(E,t,E_0,t_0)\equiv \tilde D(E,t)+\tilde D(E_0,t_0)\bigg(\frac{R_0}{R}\bigg)^4.
\end{eqnarray}
Section \ref{sec:examples} shows that non-equilibrium systems wherein $\Delta \tilde C\ne 0$ exhibit periodic fluctuations in the distribution function which damp out progressively as the system approaches an equilibrium state. 

Notice that $p_c$ plays the role of a transition probability between the states $t_0$ and $t_0+\tau'$. Appendix B  
discusses how transition probabilities can be used to describe the evolutionary path between time intervals of arbitrary length through the construction of Markov chains. 

\subsection{The Fokker-Planck approximation}\label{sec:fp}
%The previous Section derives $N(E,\{\alpha\}_\nu,t)$ from the adiabatic distribution $N(E_a,\{\alpha\}_\nu,t_0)$ under the implicit assumption that the change of orbital energy $\Delta=E-E_a$ happens instantaneously. However, it is helpful to consider the case where the invariant distribution is the end product of a process that occurs during a short, but non-zero, time interval $\tau'$. To develop this argument further we introduce a (fictitious) time-dependence in the invariant distribution, which now becomes $N(I,\{\alpha\}_\nu,t)$. In this way we can obtain an approximate solution to Equation~(\ref{eq:N1}) by taking the limit $\tau'\to 0$ of the probability functions obtained in Section~\ref{sec:diff}.
This Section relates $N(E,\{\alpha\}_\nu,t)$ to the distribution obtained in the adiabatic limit, $N_a(E_a,\{\alpha\}_\nu,t)$ by finding an approximate solution to Equation~(\ref{eq:Nt1}). To this end we apply a technique that is commonly used in variational calculus, where one assumes that 
the transformation $N_a\mapsto N$ takes place during a short, but non-zero, time interval $t_0,t_0+\tau'$. The relation between the two functions at an arbitraty time $t_0$ is made by Taylor expanding $N$ and taking the limit $\tau'\to 0$.

Recall that in a slowly-varying gravitational potential the Green function $p(E,t|E_a,t_0)$ will be sharply peaked around $E=E_a$, which calls for using $\Delta = E-E_a$ as the integration variable in Equation~(\ref{eq:Nt1}). Since $\Delta$ is given as a function of phase-space coordinates one needs to replace the probability function $p(E,t|E_a,t_0)$ by $\varphi(\Delta|E_a)$. In addition, given that $\{\alpha\}_\nu$ are assumed to be $\nu$-constants of the motion we can simplify our notation by marginalizing over $\int \d^\nu\alpha$ on both sides of Equation~(\ref{eq:Nt1}), which then becomes
\begin{eqnarray}
\label{eq:N2}
N(E,t_0+\tau')=\int \varphi(\Delta|E-\Delta)N_a(E-\Delta,t_0)\d \Delta.
\end{eqnarray} 
For an arbitrarily small $\tau'$ one can now follow the exact same steps as in \S\ref{sec:diff}. We expand the left-side of Equation~(\ref{eq:N2}) as a Taylor series in $\tau'$ up to the first order, and similarly expand the function $\varphi N$ within the right-hand integral as a Taylor series in $\Delta$ to the second order, which yields
\begin{eqnarray}
\label{eq:N3}
N_a(E,t_0)+\tau'\frac{\partial N}{\partial t}+...
= \int \d\Delta \big[\varphi- \nabla \varphi\Delta
+\frac{1}{2}\nabla^2\varphi\Delta^2 + ...]  \\ \nonumber
\times
\big[N_a 
- \nabla N_a\Delta +\frac{1}{2}\nabla^2 N_a\Delta^2 +...\big]\\ \nonumber
\simeq \int \d\Delta\big[\varphi N_a-\nabla(\varphi N_a)\Delta +\frac{1}{2}\nabla^2(\varphi N_a)\Delta^2\big]\\ \nonumber
=N_a\int \varphi\d\Delta-\nabla\bigg[N_a\int \varphi \Delta\d\Delta \bigg] 
+ \frac{1}{2}\nabla^2\bigg[N_a \int \varphi \Delta^2\d\Delta \bigg],
\end{eqnarray}
where we have defined the operator $\nabla\equiv \partial/\partial E$ and used the initial conditions $N(E,t_0)=N_a(E,t_0)$.

From Equations~(\ref{eq:norm}),~(\ref{eq:C}) and~(\ref{eq:D}) one has that $\int \varphi\d\Delta=1$, $\int \varphi\Delta \d\Delta=-\tilde C$ and $\int \varphi\Delta^2 \d\Delta=2\tilde D$, respectively. Hence, after eliminating $N_a(E,t_0)$ from both sides of Equation~(\ref{eq:N3}) we find 
\begin{eqnarray}
\label{eq:N4}
\tau'\frac{\partial N}{\partial t}\approx \nabla(\tilde C N_a)+\nabla^2(\tilde D N_a),
\end{eqnarray}
which corresponds to the Fokker-Planck equation formulated in the energy space (e.g. Spitzer 1987). 

Given that the right-hand side of Equation~(\ref{eq:N4}) is independent of the value of $\tau'$ (see discussion in \S\ref{sec:diff}) the time interval can be made arbitrarily small. By taking the limit $\tau'\to 0$ we can write the time derivative as
$$\frac{\partial N}{\partial t}\simeq \frac{N(E,t_0)-N_a(E,t_0)}{\tau'},$$
which reduces Equation~(\ref{eq:N4}) to 
\begin{eqnarray}
\label{eq:N5}
%N(E,t_0)\simeq N_a(E,t_0)+\nabla[\tilde C(E,t_0) N_a(E,t_0)]+\nabla^2[\tilde D(E,t)N_a(E,t_0)].
N\simeq N_a+\nabla(\tilde C N_a)+\nabla^2(\tilde DN_a).
\end{eqnarray}
Equation~(\ref{eq:N5}) therefore provides an approximate solution to~(\ref{eq:Nt1}) after marginalizing over $\d^\nu\alpha$. A few points of interest are worth discussing. First, in potentials that evolve adiabatically the coefficients become $\tilde C=\tilde D\simeq 0$ (see \S\ref{sec:diff}), which leads to $N\simeq N_a$. This recovers the well-known fact that in slowly-varying systems the shape of distribution function $f=N/w$ is an adiabatic invariant.
%\footnote{In thermodynamical terms the adiabatic limit is reached when the entropy associated with the probability function $N(E,\{\alpha\}_\nu,t)$ stays constant. Thermodynamic processes that obey this condition are {\it reversible}.}. 
%, which recovers the well-known fact that  function expressed in terms of adiabatic\footnote{In the adiabatic limit the entropy associated to $N(E,\{\alpha\}_\nu,t)$ remains unchanged. Thermodynamic processes that obey this condition are {\it reversible}.} invariants is also an adiabatic invariant. 
%Second, in scale-free potentials the function $N(E_a,t)$ can be expressed as a function of the initial energy $E_0$ in a straightforward way. 
%Hence, $N(E_a[E_0],t)=R^2(t)N(E_0,t_0)$, which has the same effect as changing of the energy unit by a factor $R^{-2}$.
Second, notice that the first- and second-order corrections beyond the adiabatic limit describe a net flux of particles in the energy space. From Equation~(\ref{eq:N4}) the probability flux crossing the energy layer $E$ can be estimated as
\begin{eqnarray}
\label{eq:J}
J(E)=\int_{E}\frac{\partial N(E',t)}{\partial t}\d E' \approx   C N_a + \nabla(D N_a).
\end{eqnarray}
Not surprisingly, the flux diverges in the limit $\tau'\to 0$ because in that limit the energy change becomes instantaneous. However, the bulk displacement of particles in energy space,
$\tilde J\equiv J\tau'=\tilde C N_a + \nabla (\tilde D N_a)$, is again independent of the choice of $\tau'$.

In scale-free fields it is possible to derive a straightfoward interpretation of Equation~(\ref{eq:J}). At first order the flux scales as $\tilde J\sim \tilde C N_a\propto (\dot R/R)\langle {\bf r}\cdot{\bf v} \rangle N_a$, with the brackets denoting average over microcanonical ensembles with energy $E$ (see Appendix~\ref{sec:aver}). The flux, therefore, is the result of internal macroscopic motions and oscillates in phase with the net radial velocity of the particle ensemble. If the system is in a state of quasi-dynamical equilibrium then $\langle {\bf r}\cdot{\bf v} \rangle\approx 0$, and therefore $\tilde C\approx 0$. In this limit the flux becomes diffusive, $\tilde J\approx  \nabla (\tilde D N_a)$. The fact that $\tilde D\ge 0$, $\nabla \tilde D\ge 0$ and $|E|\propto T$ (see Table~1) implies that particles will diffuse in a direction opposite to the temperature gradient. In gravitationally-bound systems this generally leads to a net particle flux from regions of high density/temperature toward regions of low density/temperature. 

\begin{figure*}
\includegraphics[width=160mm]{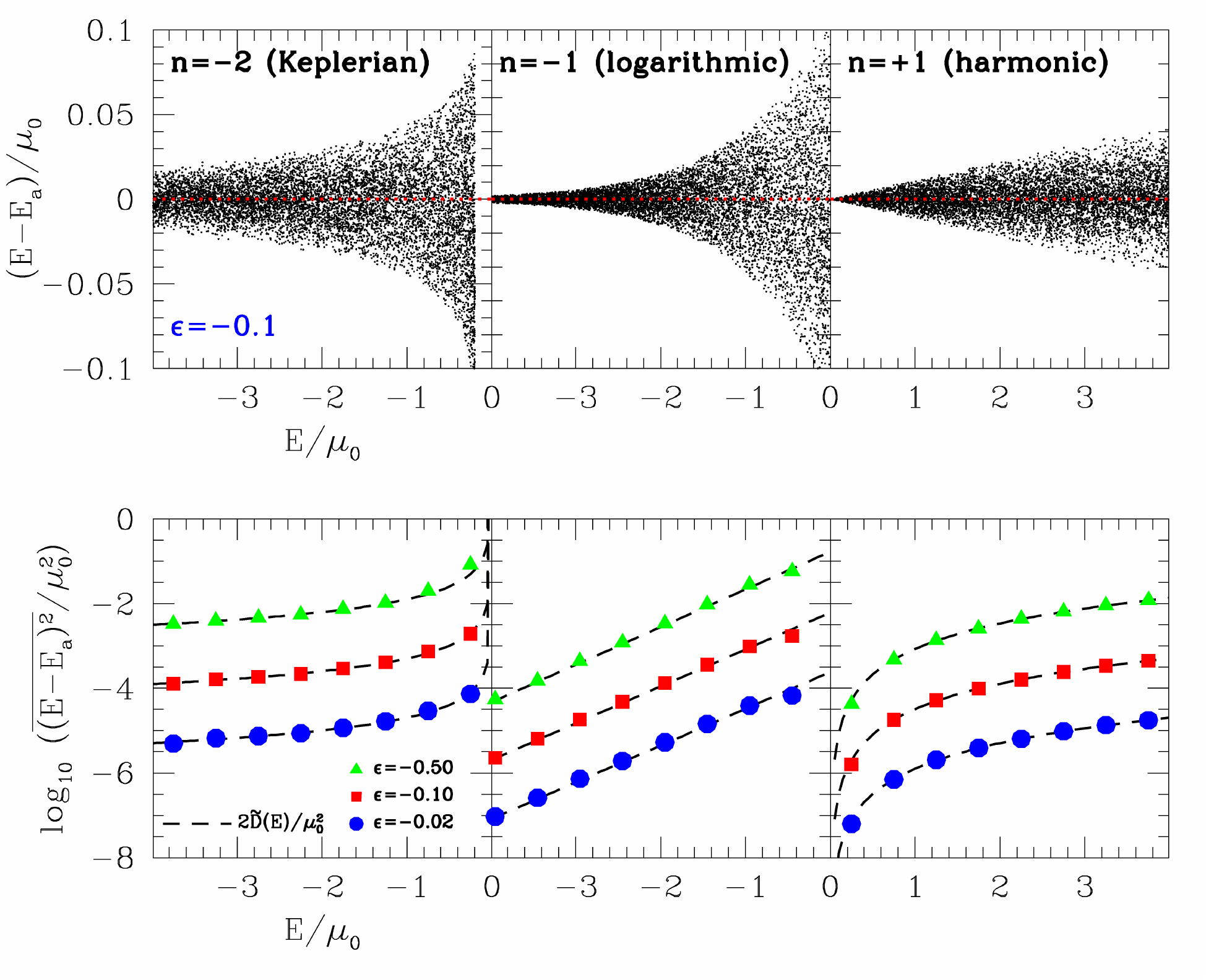}
\caption{{\it Upper panel:} Energy change with respect to the adiabatic value of $N_\star=10^4$ particles orbiting in a scale-free gravitational potential~(\ref{eq:plawp}) that grows linearly as $\mu(t)/\mu_0= 1 + \epsilon (t-t_0)/P_0$, where $\mu_0=\mu(t_0)$ and $P_0$ is the radial period of an orbit with energy $r_m=0.1 r_{\rm lim}$ at $t=t_0$ (see text). These models are run from $t=t_0$ to $t=t_0+P_0$ with $\epsilon=-0.1$.
All models show a mean energy $\tilde C=\overline{E-E_a}\approx 0$, which indicates that their evolution proceeds in quasi-dynamical equilibrium. 
In these three cases the energy dispersion increases as one moves away from the centre of the potential (i.e. toward the right-hand side of each panel). {\it Lower panel}: Energy variance as a function of energy for particles orbiting in the potentials of the upper panels. Different symbols correspond to potentials evolving at different rates. Note that the formula $\overline{(E-E_a)^2}= \tilde C^2 + 2\tilde D \approx 2\tilde D$, with $\tilde D$ derived from Equation~(\ref{eq:D3}) and Table~1 (dashed lines), is in excellent agreement with the numerical results.}
\label{fig:diff2}
\end{figure*}

\begin{figure*}  
\begin{center}
\includegraphics[width=170mm]{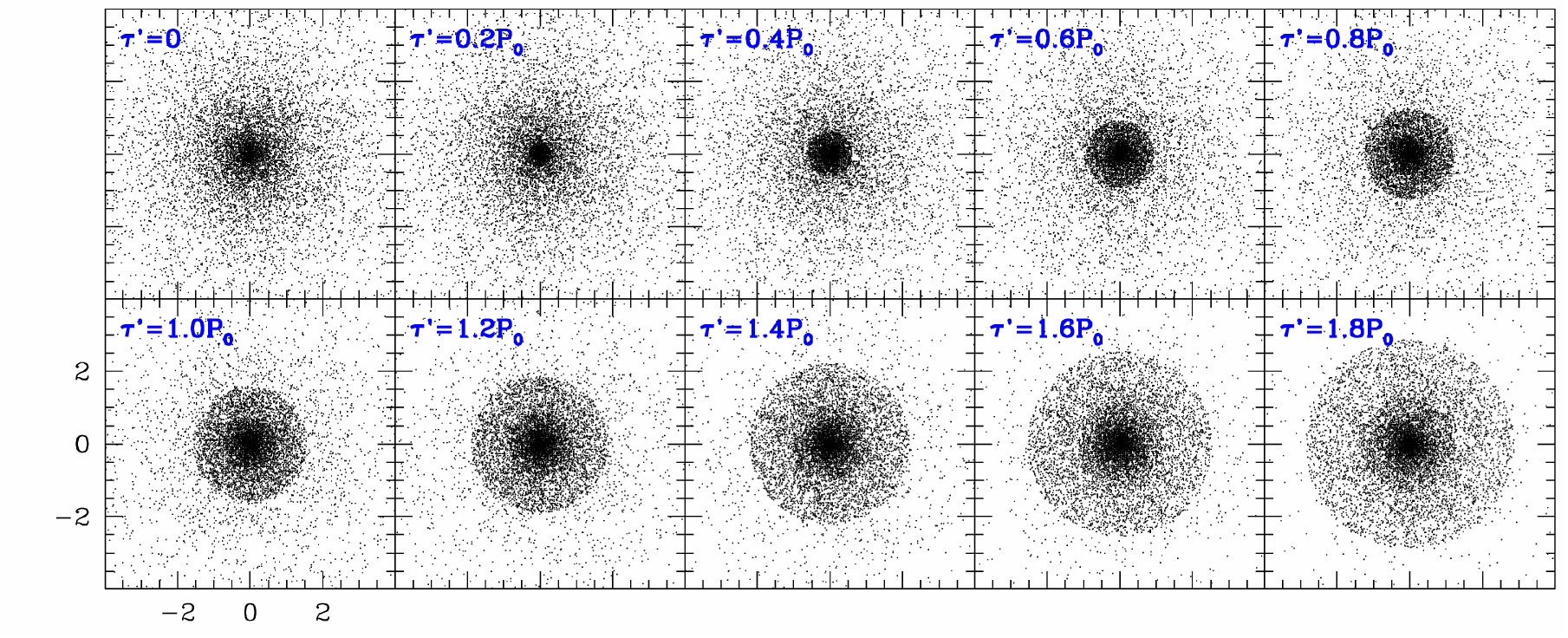}
\end{center}
\caption{Cold collapse of a model with $|2T/W|=0.04$ at $t=t_0$. Particles move in a time-dependent potential that varies at a constant rate $\epsilon=-0.1$. The system as a whole evolves toward an equilibrium configuration following a `violent relaxation' process.}
\label{fig:xy}
\end{figure*}

\begin{figure*}  
\begin{center}
\includegraphics[width=175mm]{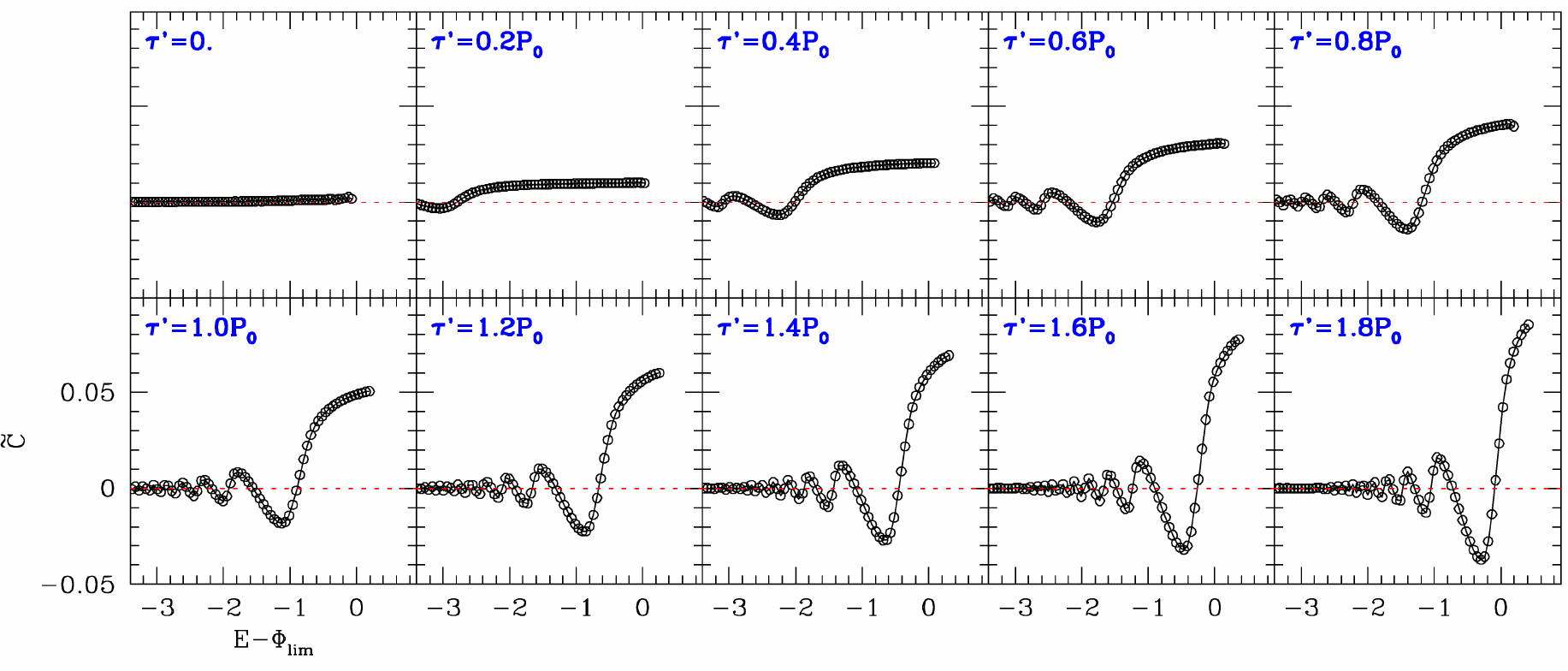}
\end{center}
\caption{Drift coefficient $\tilde C=-(\dot R/R)\langle {\bf r}\cdot{\bf v}\rangle$ as a function of energy of the model shown in Fig.~\ref{fig:xy}. Notice that at $t=t_0$ the system is phased mixed (i.e. $\langle {\bf r}\cdot{\bf v}\rangle=0$) but out of virial equilibrium, $|2T/W|=0.04$. By the end of the simulation only the internal regions of the potential have reached a state of quasi-dynamical equilibrium ($\tilde C\approx 0$).}
\label{fig:C}
\end{figure*}

\section{$N$-body experiments}\label{sec:examples}
In this Section we run a number of restricted $N$-body experiments that illustrate the evolution of tracer particles orbiting in the time-dependent scale-free potential given by Equation~(\ref{eq:plawp}). 
To simplify the analysis we shall consider spherical power-law potentials that evolve linearly with time as
\begin{equation}
\label{eq:mu}
\frac{\mu(t)}{\mu_0}=1+\epsilon\frac{t-t_0}{P_0},
\end{equation}
where $P_0=P(E,t_0)=2\int_0^{r_m(E)}\d r/\sqrt{2(E-\Phi[r,t_0])}$ is the radial period of an orbit with $r_m=\mu_0=1$ at $t=t_0$. 

Particles are distributed within a distance range from the centre of the potential, $r_{\rm min} \le r\le r_{\rm lim}$, with $r_{\rm min}=0.1$ and $r_{\rm lim}=10$.
The initial radii and velocities of the particle ensemble are drawn from an initial distribution function $f(E_0,t_0)$ using a rejection algorithm, while the position and velocity vectors are isotropically distributed over a sphere.
The equations of motion~(\ref{eq:eqmot}) of each individual particle are then integrated forward in time using a leap-frog algorithm whose time-step is chosen such that the energy is conserved at least at a $10^{-4}$ level in the static case ($\epsilon=0$).

\subsection{Analytical diffusion coefficients}\label{sec:diffnum}
First we shall study a few examples where the coefficients $\tilde C$ and $\tilde D$ can be derived analytically using the dynamical invariants introduced in \S\ref{sec:inv}.

Let $f(E_0,t_0)$ be a distribution function where the number of particles in the energy interval $E,E+\d E$ has a fixed value independently of $E$. From Appendix~\ref{sec:aver} this condition implies
\begin{equation}
\label{eq:fcons}
f(E_0,t_0)=\frac{N_0}{\omega(E_0,t_0)}.
\end{equation}
Integrating on both sides of Equation~(\ref{eq:fcons}) over the phase-space volume $\d^6\Omega$ shows that the total number of particles in the model is $N_0 \Delta E$, where $\Delta E$ corresponds to the range of energies covered by the particle ensemble.

Fig.~\ref{fig:diff2} shows the variation of energy relative to the adiabatic value, i.e. $E-E_a(t)$, as a function of the orbital energy in time-dependent potentials with a growth rate $\epsilon=-0.1$.
The left, middle and right panels adopt Keplerian ($n=-2$), logarithmic ($n=-1$) and harmonic ($n=+1$) potentials, respectively. In general, we find that the displacements from the adiabatic limit are on average very small, $\overline{E-E_a}\approx 0$, which indicates that the $N$-body models are initially in dynamical equilibrium. In contrast, the variance $\overline{(E-E_a)^2}$ increases significantly as one moves away from the centre of the potential (i.e. toward the right-hand side of each panel). This behaviour can be easily understood using the diffusion coefficients derived in \S\ref{sec:coeffplaw}. In a scale-free gravitational field $\overline{(E-E_a)^2}=2\tilde D\propto r_m^{3+n}$, where $r_m(E,t)$ is the maximum radius that a particle with an energy $E$ can reach in the potential~(\ref{eq:plawp}), that is 
\begin{equation}
\label{eq:rm}
r_m(E,t)=
\begin{cases}
\big[(n+1) (E-\Phi_\infty)/\mu(t)\big]^{1/(1+n)} & ,n\ne - 1 \\ 
r_{\rm lim}\exp[E/\mu(t)] & , n=-1.
\end{cases}
\end{equation}
The constant in front of the potential~(\ref{eq:plawp}) is $\Phi_\infty=0$ for the Keplerian and harmonic cases, and $\Phi_\infty=-\Phi(r_{\rm lim},t_0)$ for the logarithmic one.
Hence, from Equation~(\ref{eq:rm}) it follows that $\tilde D\propto E^{-1}$, $\tilde D\propto e^{2E/\mu(t)}$ and $\tilde D\propto E^{2}$ for power-law indices $n=-2, -1,$ and $+1$, respectively. The lower panels of Fig.~\ref{fig:diff2} shows that the diffusion coefficients given by Equation~(\ref{eq:D3}) and listed in Table~1 (dashed lines) provide an excellent match against the $N$-body models (symbols) over a large range of energies and growth rates.

\begin{figure*}
\includegraphics[width=176mm]{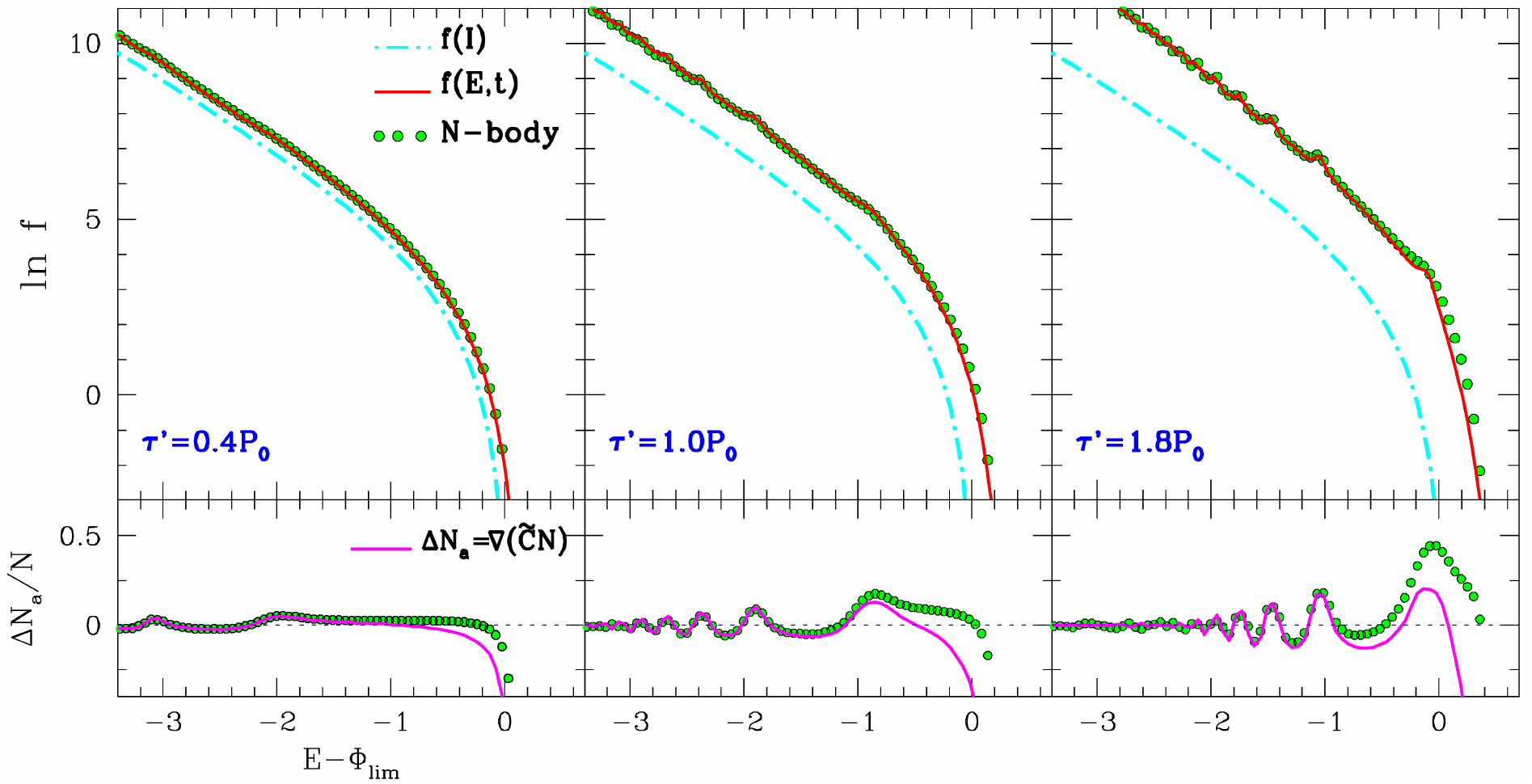}
\caption{{\it Upper panels:} Distribution function $f=N/\omega$ of the models shown in Fig.~\ref{fig:xy} at three different snap-shots. Red solid lines correspond to the Green convolution given by Equation~(\ref{eq:Nt2}) with coefficients $\tilde C(E,t)=-(\dot R/R)\langle {\bf r}\cdot{\bf v}\rangle$ and $\tilde D(E,t)=(\dot R/R)^2\langle ({\bf r}\cdot{\bf v})^2\rangle$ measured from the phase-space locations of the $N$-body particles. As expected, the distribution $f(I)$ of Equation~(\ref{eq:N1}) (cyan dashed curves) remains invariant throughout the evolution of the system. {\it Lower panels:} Deviation of the $N$-body models from the adiabatic distribution, $\Delta N_a\equiv N(E,t)-N_a(E,t)$ (green dots). The Fokker-Planck approximation, Equation~(\ref{eq:N5}), is shown with magenta solid lines. Note that the ripples and troughs of the curves are located at energies where the gradient $\nabla(\tilde C N)$ finds local maxima and minima, respectively. Departures from the adiabatic solution are particularly strong in the outskirts of the system, $E\gtrsim \Phi_{\rm lim}$.}
\label{fig:dist_FP}
\end{figure*}
\subsection{Violent relaxation}\label{sec:dfnum}
In this Section we study the evolution of $N$-body models with a virial ratio $|2T/W|\ll 1$ at $t=t_0$. Setting the initial kinetic pressure to such a low value leads to a rapid infall of particles toward the centre of the potential, a process typically known as {\it cold collapse}. Given that the potential varies as a function of time the orbital energies of individual particles change in a non-trivial way as the system approaches equilibrium (Lynden-Bell 1967; Mo, van den Bosch \& White 2010).

As an illustration of this process we generate $N$-body equilibrium realizations of a lowered Maxwellian distribution in a logarithmic potential
\begin{equation}
\label{eq:df0}
f(E_0,t_0)=
\begin{cases}
A\big[e^{\frac{-E_0+\Phi_{\rm lim}}{\sigma^2}}-1\big] &, E_0<\Phi_{\rm lim},\\
0 & ,E_0\ge\Phi_{\rm lim},
\end{cases}
\end{equation}
where $A$ is a normalization factor, and $\Phi_{\rm lim}=\mu_0\ln(r_{\rm lim})$ is an energy truncation. To simplify our models we set 
$\sigma^2=\mu_0/2=3 T$, where $T$ is the kinetic energy associated with the potential (see Table~1). 

Next, we multiply the velocities of all particles by a factor $q$. To highlight non-equilibrium features we choose a small value, $q=0.2$, which leads to $|2T/W|=q^2=0.04$. Such a low virial ratio guarantees the collapse of the system on a time-scale comparable to its free-fall time. 
Fig.~\ref{fig:xy} shows ten snap-shots of a model orbiting in a time-dependent potential that evolves at a rate $\epsilon=-0.1$. Cold collapse happens early on ($\tau'\approx 0.2 P_0$) and leads to the formation of shell structures that move progressively towards larger radii with time. 
By the end of the simulation, $\tau'\approx 2.0 P_0$, the model has not yet reached dynamical equilibrium. By definition the system is in a state of `incomplete relaxation'\footnote{These models provide a useful representation of the dynamical state of the outer regions of galactic haloes and galaxy clusters, where dynamical times are comparable to the age of the Universe and phase mixing becomes very inefficient (e.g. H{\'e}non 1964, Lynden-Bell 1967; Dehnen 2005).}.

Fig.~\ref{fig:C} shows the drift coefficient $\tilde C=-(\dot R/R)\langle {\bf r}\cdot{\bf v}\rangle$ as a function of energy for the snap-shots shown in Fig.~\ref{fig:xy}. The initial $N$-body models are phase mixed, such that $\tilde C\propto \langle {\bf r}\cdot{\bf v}\rangle\approx 0$ for all energies. However, the fact that $|2T/W|\ll 1$ implies that the model is far from a virialized state. 
As the system begins to collapse the averaged radial velocity of the particle ensemble is negative at all radii and $\partial \tilde C/\partial t> 0$ at all energies. 
At slightly later times, $\tau'\gtrsim 0.2 P_0$, the coefficient $\tilde C$ begins to exhibit coherent fluctuations in the inner-most regions of the potential (left side of the panels), as particles with short orbital period go through pericentre and start moving toward larger radii. In the outskirts, however, particles are still falling in from large distances, which translates into positive values of $\tilde C$. The negative crests are associated with the shell features of Fig.~\ref{fig:xy}. 
Given that in a potential with $n=-1$ the orbital period decreases toward the central regions of the potential as
%\begin{equation}\label{eq:P}
$$P(E,t)=(2\pi/\mu)^{1/2}r_m=(2\pi/\mu)^{1/2}r_{\rm lim}\exp(E/\mu),$$
%\end{equation}
fluctuations in $\tilde C$ damp out progressively from inside out. By $\tau'=1.9 P_0$ the mean radial velocity of particles with $E-\Phi_{\rm lim}\lesssim -2$ is $\langle {\bf r}\cdot{\bf v}\rangle\approx 0$, signalling that the inner regions of the system are phase-mixed and evolving in a state of quasi-dynamical equilibrium.

Fig.~\ref{fig:dist_FP} illustrates the complexity involved in describing the state of systems undergoing violent relaxation. Green dots show the distribution function of the $N$-body models plotted in Fig.~\ref{fig:xy} at three different snap-shots. The decreasing potential ($\epsilon<0$) shifts the orbital energies to higher values, which leads to a non-monotonic increase of $f(E,t)$ at fixed energies. Interestingly, the distribution function is not completely smooth. Non-equilibrium features arise in the inner-most regions of the potential and propagate toward high energies as time goes by. After constructing a larger suite of $N$-body models (not shown here) we find that the amplitude of these fluctuations increases for larger $\epsilon$ and smaller $q$ values, which correspond to faster growth rates and higher radial anisotropies, respectively.

Equation~(\ref{eq:Nt2}) (red solid lines) is able to capture these complexities and provide an accurate statistical description of the non-equilibrium state of the system. For simplicity we assume that the transition probability $p_c(E,t|E_0,t_0)$ is independent of the evolutionary path taken by the system between $t_0$ and $t_0+\tau'$ (see Appendix B).
The Green convolution is solved by setting $\tilde C(E_0,t_0)=0$ and computing $\tilde D(E_0,t_0)$ analytically from Equation~(\ref{eq:D3}) and Table~1. The coefficients $\tilde C(E,t)$ and $\tilde D(E,t)$ are measured from the phase-space coordinates of the $N$-body particles at $t=t_0+\tau'$ as $\tilde C(E,t)=-(\dot R/R)\langle {\bf r}\cdot{\bf v}\rangle$ and $\tilde D(E,t)=(\dot R/R)^2\langle ({\bf r}\cdot{\bf v})^2\rangle$. Note that in this case the solutions to Equation~(\ref{eq:Nt2}) are not purely statistical, as our calculation of the coefficients adds additional dynamical information. 

The Fokker-Planck approximation provides useful insight into the dynamical mechanisms that drive the shape of the distribution function.
The lower panels of Fig.~\ref{fig:dist_FP} plots the difference between the $N$-body distribution function and the adiabatic solution given by Equation~(\ref{eq:Nta}) at fixed energy values (green dots). From Equation~(\ref{eq:mu}) particles respond adiabatically to a time-varying force if their orbital periods obey $P (\dot \Phi/\Phi_0)= \epsilon (P/P_0) \ll 1$. 
Since $P$ increases exponentially with the particle energy we find the distribution $N$ evolves adiabatically ($\Delta N_a\approx 0$) at $E\ll \Phi_{\rm lim}$, while strong departures from the adiabatic solution ($|\Delta N_a|\sim N$) are visible at $E\gtrsim\Phi_{\rm lim}$, where the potential changes significantly during an orbital period, i.e. $P (\dot \Phi/\Phi_0) \sim 1$. As a result, the Fokker-Planck approximation (solid magenta lines) becomes less accurate in the outskirts of the system. Note that similar deviations are also visible in the upper panels at $\tau'=1.8 P_0$, albeit with a lesser magnitude. 

The existence of internal macroscopic motions ($\tilde C\ne 0$) leads to fluctuations of the distribution function which travel toward high energies, as shown in Fig.~\ref{fig:C}. 
Comparison with the solid magenta lines shows that the ripples and troughs of the fluctuations are located at energies where the gradient $\nabla(\tilde C N)\equiv \partial (\tilde C N)/\partial E$ finds local maxima and minima, respectively. Notice that by the end of the simulation ($\tau'=1.8 P_0$) relaxation is still `incomplete'.

In phase-mixed regions relaxation proceeds in quasi-dynamical equilibrium ($\tilde C\approx 0$), and the evolution of the distribution function becomes diffusive, as discussed in \S\ref{sec:fp}. The {\it final} state that emerges from violent relaxation can be found by setting $\tilde C(E,t)=0$ in Equation~(\ref{eq:Nt2}).

\section{Summary \& discussion }\label{sec:dis}
This paper introduces a probability theory that describes the non-equilibrium evolution of large particle ensembles orbiting in a time-dependent gravitational field. The theory is constructed on the basis of dynamical invariants (quantities conserved along the phase-space path of a particle motion) and does not rely on maximum-entropy probabilities or ergodicity assumptions.

A fundamental result of this paper states that the evolution of particles with the same orbital energy (the so-called microcanonical ensemble) has the same form as Einstein's equation for freely diffusing particles with coefficients that are closely related to virial quantities, such as the moment of inertia and the temperature of the system. 
Since the general solution to the diffusion equation corresponds to a Green function, the non-equilibrium distribution function of systems composed of an infinite number of microcanonical subsystems is found by convolving the initial distribution function with the solution to the diffusion equation. The Fokker-Planck equations~(\ref{eq:N5}) provide an approximate solution to this type of Green convolutions. Thus, we conclude that the problem of collisionless relaxation in the linear regime reduces to the derivation of diffusion coefficients. 

The analogy between our mathematical formalism and stochastic calculus is undeniable and also somewhat striking.
Indeed, at first glance it is not completely obvious why equations that are common to a large range of stochastic processes do also describe systems governed by fully deterministic equations of motion. To answer this question it is useful to examine the role of the probability function $\varphi(\Delta)$ in some detail.
In Einstein (1905) the explicit form of $\varphi(\Delta)$ contains all information on the dynamics of the collisions that drive the motion of Brownian particles. However, its role is that of a nuisance function, for the diffusion coefficients are derived indirectly from observations of macroscopic quantities (such as the mean squared displacement of the particles as well as 
the temperature and the viscosity of the medium where the particles are suspended) over an extended period of time, $\tau'$. 
The great virtue of this approach is that it permits a derivation of the probability distribution of random displacements at the time $t_0=t_0+\tau'$ from the distribution at $t=t_0$ without having to specify the form of $\varphi(\Delta)$! 
Although this frees us from the necessity to understand the mechanics of Brownian motion at a microscopic level, it becomes necessary to assume that time-averaged properties emerge from phase-space averages of microscopic quantities, which constitutes the crux of the {\it ergodic principle}. 
%Yet, there is a price to pay for freeing us from the necessity of understanding the dynamical mechanisms behind the Brownian motion, as it becomes necessary to assume that time-averaged properties of the particle ensemble emerge from phase-space averages of microscopic quantities, which is typically known as the {\it ergodic principle}. 
Unfortunately, the equivalence between time average and average over ensembles only arises when the system can visit all the possible microstates, many times, during a long period of time, which in general cannot be guaranteed in non-equilibrium systems. 
In contrast, in the present work the form of $\varphi(\Delta)$ is known and corresponds to the microcanonical distribution function~(\ref{eq:micro}), while the dynamical invariants of Section~\ref{sec:inv} define the phase-space dependence of $\Delta$.
%This allows us calculate averages of $\Delta$ over the phase-space volume accessible to the particles that form the microcanonical ensemble. 
In systems with known dynamical invariants it is therefore possible to find {\it self-consistent} solutions to the diffusion equation without resorting to time averages of macroscopic quantities, thus removing the dependence on the time interval $\tau'$ from the solution and -- more importantly -- the necessity to rely on ergodicity assumptions in order to describe the evolution of gravitating systems. 

Alas, not all potentials admit analytical solutions to the diffusion equation. For example, in systems that are not initially virialized a derivation of the diffusion coefficients requires precise knowledge of the trajectories of individual particles in phase space, rendering the problem analytically intractable. Also, to this date an analytical derivation of energy invariants is only feasible in scale-free gravitational potentials (see P13). And yet the theory has many interesting applications that go beyond the analytical cases explored here. For example notice that in this paper the time evolution of the potential has been intentionally detached from the spatial distribution of the particles under scrutiny. In essence, we have treated particles as mass-less tracers of an underlying gravitational field which is completely specified by the particle coordinates and the time (that is, the field is purely mechanical and not a statistical object). Although there is a large number of astrophysical objects that meet these conditions (stars in dwarf spheroidal galaxies, tidal streams in galactic haloes, and galaxies in galaxy clusters are a few prototypical examples of kinematic tracers in time-dependent potentials), an exciting task for the future may be to study a more constrained case where the initial state of the system is specified by $f({\bf r},{\bf v},t_0)$ at $t=t_0$, and the evolution of particle ensemble is driven by its own self-gravity 
$$ \nabla^2\Phi({\bf r},t)=4\pi G \int f({\bf r},{\bf v},t)\d^3 v,$$
with $f=N/\omega$ given by Equation~(\ref{eq:N1}). A solution to this problem will provide a deep insight into the evolution and stability of self-gravitating collisionless systems.
%For example, this is the problem we need to solve in order to follow the evolution of gravitationally-bound dark matter haloes from a close-to-homogeneous and isotropic matter distribution that is expanding with time.

By considering time-dependent potentials with a fixed spatial symmetry we have confined the diffusion process outlined in \S\ref{sec:diff} to the energy dimension. Yet, our formalism can be straightforwardly extended to systems with a varying spatial symmetry. As an illustration, let us consider a system in which both, the angular momentum and the energy of individual particles are allowed to vary with time. The Green function appearing in Equation~(\ref{eq:N1}) has now one extra dimension, i.e. $p(E,L,t|E',L',t_0)$, which defines the probability that a particle with an energy $E'$ and angular momentum $L'$ at the time $t=t_0$ has an energy in the interval $E,E+\d E$ and an angular momentum in the interval $L,L+\d L$ at the time $t=t_0+\tau'$. 
Armed with the new probability function one can follow the same steps as in \S\ref{sec:fp} to derive the two-dimensional Fokker-Planck equations
\begin{eqnarray}
\tau'\frac{\partial N}{\partial t}\approx -\frac{\partial }{\partial E}[\overline{\Delta} N] - \frac{\partial }{\partial L}[\overline{\Delta L} N]+ \frac{1}{2}\frac{\partial^2 }{\partial E^2}[ \overline{\Delta^2 }N]  \\ \nonumber
+\frac{1}{2}\frac{\partial^2 }{\partial L^2}[\overline{\Delta L^2} N] + \frac{\partial^2 }{\partial E \partial L}[\overline{\Delta (\Delta L)}N],\nonumber
\end{eqnarray}
where the coefficients correspond to phase-space averages of $\Delta L\equiv L-L'$, $\Delta L^2=(L-L')^2$ and $\Delta(\Delta L)=(E-E')(L-L')$, as indicated by Equation~(\ref{eq:xal}).
%\begin{eqnarray}\nonumber
%\overline{\Delta L}(L,E,t)= \int (\Delta L)\eta(\Delta L,L,E) \d (\Delta L)\\ \nonumber
%\overline{(\Delta L)^2}(L,E,t)= \frac{1}{2}\int (\Delta L)^2\eta(\Delta L,L,E) \d (\Delta L)\\ \nonumber
%\overline{\Delta (\Delta L)}(L,E,t)=\int  \varphi(\Delta,E) \Delta \d \Delta\int(\Delta L)  \eta(\Delta L,L,E) \d (\Delta L). \nonumber
%\end{eqnarray}
The simplest case corresponds to an evolutionary process in which the change of angular momentum is independent of the orbital energy of individual particles. The cross-coefficient becomes $\overline{\Delta (\Delta L)}=0$, which leads to a Green function $p(E,L,t|E',L',t_0)$ with a relatively simple form
$$\frac{1}{ [(4\pi)^2\tilde D\tilde D_L]^{1/2}}\exp\bigg[-\frac{(\Delta +\tilde C)^2}{4 \tilde D} -\frac{(\Delta L +\tilde C_L)^2}{4 \tilde D_L}\bigg],$$
where $\overline{\Delta L}=-\tilde C_L(L',t)$ and $\overline{\Delta L^2}-\overline{\Delta L}^2=2\tilde D_L(L',t)$ play the same role as $\tilde C$ and $\tilde D$ in the energy space. In general, radial orbit instabilities (see H\'enon 1973, Huss et al. 1999, MacMillan et al. 2006) will lead to $\overline{\Delta (\Delta L)}\ne 0$ even in spherical potentials. The importance of orbital inestabilities has been recently highlighted by Pontzen et al. (2015), who argue that collisionless systems evolve toward attractor states which are maximally stable when external perturbations are applied. However, it is important to emphasize that most galaxies never reach a fully-relaxed state within a Hubble time, which may be the reason for the observed mismatch between the distribution function derived from maximum-entropy arguments and those observed in cosmological $N$-body simulations (Pontzen \& Governato 2013). Given that our theory can describe the non-equilibrium state of self-gravitating systems, a derivation of the crossed-coefficients $\overline{\Delta (\Delta L)}$ will shed light on the way in which the distribution function actually evolves towards its maximally-stable limit.

%\footnote{In general, radial orbit instabilities will lead to $\overline{\Delta (\Delta L)}\ne 0$ even in spherical potentials (e.g. Pontzen et al. 2015 and references therein).}.
%$\overline{\Delta L}(L,E_a,t)= \int (\Delta L)\eta(\Delta L,L,E_a) \d (\Delta L)$, $\overline{(\Delta L)^2}(L,E_a,t)\equiv \int (\Delta L)^2\eta(\Delta L,L,E_a) \d (\Delta L)$ and $\overline{\Delta (\Delta L)}(L,E_a,t)\equiv \int  \Delta p(E_a,t) \d \Delta\int(\Delta L)  \eta(\Delta L,L,E_a) \d (\Delta L)$.

Finally, the fact that the evolution of the microcanonical distribution has the same form as Einstein's equation for purely stochastic processes offers the tantalizing possibility to incorporate the effects of random particle-particle collisions into our probability theory in a natural way. Such an extension may be useful, for example, to study a range of dynamical processes taking place in planetary systems, dense stellar objects and/or self-interacting dark matter (SIDM) haloes.

\section{Acknowledgements}
I would like to thank Justin Read and Frank van den Bosch for their insightful questions, which proved to be key to clarify essential aspects of the theory.
Also, many thanks to Matt Walker for pointing me to Donald's paper on Dirac's cosmology, and to the DwaIN network for t\^et\^e-\`a-t\^et\^e discussions.

{}

\appendix

\section{Statistical averaging}\label{sec:aver}
The large number of particles typically enclosed in astrophysical systems, together with the necessary incompleteness of observational data, makes phase-space averaging one of the most useful tools in Statistical Mechanics. Here we discuss the use of a microcanonical distribution to calculate phase-space averages of macroscopic quantities. 

 In order to express the distribution function introduced in \S\ref{sec:df} as a function of phase-space coordinates one needs to map points in the integral-of-motion space onto the phase-space volume $\d^6\Omega=\d^3{\bf r}\d^3{\bf v}$ through the equation
\begin{eqnarray}
\label{eq:intspace}
f({\bf r},{\bf v},t)\d^6 \Omega &=& f(E,\{\alpha\}_\nu,t)\bigg|\frac{\partial ({\bf r},{\bf v})}{\partial (E,\alpha_1,...,\alpha_\nu)}\bigg|\d E\d^\nu\alpha \\ \nonumber
&\equiv& N(E,\{\alpha\}_\nu,t)\d E\d^\nu\alpha,
\end{eqnarray} 
where $[w]\equiv\partial ({\bf r},{\bf v})/\partial(E,\alpha_1,...,\alpha_\nu)$ is the {\it matrix of density of states}. The Jacobian of this matrix, $w(E,\{\alpha\}_\nu,t)$, defines the maximum space-space volume that particles with a given combination of integrals of motion can potentially sample. With this definition the probability of finding a particle within the energy interval $(E,E+\d E)$ can be found by marginalizing Equation~(\ref{eq:intspace}) over $\{\alpha\}_\nu$, that is $N(E,t) = \int N(E,\{\alpha\}_\nu,t) \d^\nu\alpha$, with the probability function normalized such that $\int N(E,t)\d E=1$.

A very useful thermodynamical concept corresponds to {\it microcanonical ensembles}. Consider a gravitating system where all particles have a known orbital energy. Now isolate a subset of particles with a specific energy, $E$. Even if the integration of the equations of motion were feasible, that information alone would not be enough to determine the trajectories of those particles in phase space, for it is not possible to define the initial conditions of the problem. This means that in general we will have a complete lack of knowledge of their {\it microscopic state}.
The only available information reduces to the condition imposed on their phase-space locations, which must be confined to a hyper-surface in phase space, $\Omega(E)$, where $E=H({\bf r},{\bf v},t)$, and $H=1/2 {\bf v}^2 + \Phi$ is the Hamiltonian of the system.

A key question is, how are those particles distributed on the single-energy hypersurface $\Omega(E)$? If they are homogeneously distributed one can easily construct an {\it ensemble of microcanonical states} by randomly sampling $\Omega(E)$ a very large number of times. By construction all microstates would be equally probable, and intuition tells us that the averaged properties of a microstate would be indistinguishable from the rest within small fluctuations. But what do we mean exactly with `equally probable'? Here we can either adopt a frequentist definition, where one measures the averaged properties of each microstate at different times in order to assign a certain probability, or a statistical one, where the assigned probability is based on the density of particles on $\Omega(E)$. If the system is {\it ergodic} both probabilities are the same within small fluctuations, which immediately leads to an equivalence between time average and average over microstates. However, one can easily find examples of gravitating systems where ergodicity is broken. Consider, for example, a Globular Cluster losing stars to the host's tidal field. Initially these stars will distribute along tidal tails with a small spread in orbital energy, which can be neglected for the sake of argument. Clearly, since tidal tails do not sample $\Omega(E)$ in a randomn fashion the ergodic hypothesis would lead to a wrong description of their macroscopic properties. However, if we wait long enough we would see the tidal tails spreading throughout larger segments of this hypersurface, a process that is typically known as {\it phase mixing}. Thus, the system becomes ergodic in the limit $t\to \infty$, as the occupation of $\Omega(E)$ evolves toward a global maximum (see P13 for numerical illustrations of this process). Although one can formally construct ensembles of `unmixed' microstates (see \S\ref{sec:micro}), a description of their macroscopic properties will require dynamical information on the trajectories of particles in phase space, which in general renders the problem non analytical, as discussed in \S\ref{sec:examples}. In the remainder of the Appendix we focus on phase-mixed systems, for which microcanonical averages can be straightforwardly derived.

For spherical \& isotropic systems the density of states of a microcanonical ensemble of particles with an energy $E$ reduces to $\omega = \partial \Omega/\partial E$, which can be written as
\begin{eqnarray}
\label{eq:w}
\omega(E,t)&=&\frac{\partial }{\partial E}  \int_{\Omega(E)} \Theta(E-H)\d^6\Omega=\int \delta (E-H)\d^6\Omega\\\nonumber
&=&(4\pi)^2\int_0^{r_m}[2(E-\Phi)]^{1/2}r^2\d r,
\end{eqnarray} 
wher $\Theta(E-H)$ the Heaviside function, and $r_m(E,t)$ is the maximum radius that particles with energy $E$ can reach, that is $\Phi(r_m,t)=E$. 
The second equality in Equation~(\ref{eq:w}) follows from the relation between the delta and the Heaviside functions, that is $\delta (x-a)=\frac{\d}{\d x} \Theta(x-a)$, which also helps to simplify the calculations in Section~\ref{sec:diff}. 

Let $\langle X(E)\rangle$ be the isotropic average of an arbitrary quantity $X({\bf r},{\bf v})$ over the phase-space volume of a microcanonical ensemble of particles with energy $E$. From Equation~(\ref{eq:w}) this can be written
\begin{eqnarray}
\label{eq:xam}
\langle X(E)\rangle &=&\frac{1}{\omega} \frac{\partial }{\partial E}  \int \Theta(E-H)X({\bf r},{\bf v})\d^6\Omega \\ \nonumber
&=&\frac{1}{\omega}  \int \delta (E-H) X({\bf r},{\bf v}) \d^6 \Omega=\int f_E({\bf r},{\bf v},t) X({\bf r},{\bf v}) \d^6 \Omega.
\end{eqnarray}
where $f_E$ is the so-called the {\it microcanonical distribution function}
\begin{equation}
\label{eq:micro}
f_E({\bf r},{\bf v},t)\d^6\Omega=\frac{1}{\omega}\times\delta(E-H)\d^6\Omega,
\end{equation}
which defines the probability of finding a particle with an energy $E$ in the phase-space volume $\d^6\Omega$ centred at the coordinates $({\bf r},{\bf v})$ at the time $t$ (e.g. Landau \& Lifshitz 1980).

Averaging $X$ over all of phase space of the system yields
\begin{eqnarray}
\label{eq:xa}
\langle X\rangle&=&\int f({\bf r},{\bf v},t)  X({\bf r},{\bf v}) \d^6 \Omega \\ \nonumber
&=&\int \d E\frac{N(E,t)}{\omega}\frac{\partial}{\partial E}\int \Theta(E-H)X({\bf r},{\bf v}) \d^6 \Omega(E) \\ \nonumber
&=&\int \d E\frac{N(E,t)}{\omega}\int \delta(E-H)X({\bf r},{\bf v}) \d^6 \Omega \\ \nonumber
&=&\int \d E N (E,t)\int f_E({\bf r},{\bf v})  X({\bf r},{\bf v})  \d^6\Omega\\ \nonumber
&=&\int N (E,t)\langle X(E)\rangle\d E= \overline{X},
\end{eqnarray}
highlighting the equivalence between phase-space average, $\langle X\rangle$, and average over microcanonical ensembles, $\overline{X}$.

For spherical \& anisotropic systems the distribution function can be written as $f=f(E,L,t)$, where $L$ is the modulus of the angular momentum vector. To calculate the volume accessible to stars with energy $E$ and angular momentum $L$ it is useful to define a coordinate system such that the tangential and radial velocity components can be written as $v_t=v \sin\theta$ and $v_r=v \cos\theta$. In these coordinates the angular momentum becomes $L=r v \sin\theta$ and $\d L= r v_r\d \theta$, so that $\sin\theta\d \theta=L\d L/(r v v_r)$. The sign of the radial velocity component changes as 
the particle moves toward and away from the centre of the potential, becoming zero at the peri and apocentre radii, $r_p$ and $r_a$, respectively. With this information in mind the density of states of an anisotropic system can be written as
\begin{eqnarray}
\label{eq:wa}
\omega(E,L,t)&=&\int \delta (E-H)\delta (L-r v \sin \theta)\d^6\Omega\\ \nonumber
&=&(4\pi)^2\int_{r_p}^{r_a} r^2\d r\times v^2\d v\sin\theta\d \theta\\ \nonumber
&=&(4\pi)^2 L \int_{r_p}^{r_a}\frac{\d r}{v_r} =8\pi^2 L P(E,L,t),
\end{eqnarray} 
where $P(E,L,t)=2\int_{r_p}^{r_a} \d r/v_r$ is the period of an orbit with energy $E$ and angular momentum $L$ at the time $t$.

For anisotropic systems Equation~(\ref{eq:xa}) becomes
\begin{eqnarray}
\label{eq:xal}
\langle X\rangle&=&\int f({\bf r},{\bf v},t)  X({\bf r},{\bf v}) \d^6 \Omega  \\ \nonumber
&=&\int \d E\d L\frac{N}{\omega}\int \delta(E-H)\delta (L-r v \sin \theta)X({\bf r},{\bf v}) \d^6 \Omega \\ \nonumber
&=&\int N (E,L,t)\langle X(E,L)\rangle\d E\d L= \overline{X},
\end{eqnarray}
where $\langle X(E,L)\rangle$ denotes the average of $X$ for all stars of a particular $E$ and $L$.

\section{Markov chains}\label{sec:markov}
Equation~(\ref{eq:Nt2}) shows that $p_c(E,t|E_0,t_0)$ plays the role of a transition probability between the states $t=t_0$ and $t=t_0+\tau'$, {\it where $\tau'$ is a time interval of arbitrary length}. As such,  
$p_c$ only depends on the initial and final states of the system and not on the steps taken to get there, a remarkable property which is worth discussing in some depth.

Consider an evolutionary path that goes from $t_0$ to $t_0+\tau'$ through an intermediate step $t_0< t_1\le t_0+\tau'$. How would the results obtained in \S\ref{sec:timedf} change? As a first step we notice that 
since $N(I,\{\alpha\}_\nu$) is invariant each individual state must obey Equation~(\ref{eq:N1}), hence
\begin{eqnarray}\label{eq:tinterm}
\int p(I|E_0,t_0)N(E_0,\{\alpha\}_\nu,t_0)\d E_0 = \int p(I|E_1,t_1)N(E_1,\{\alpha\}_\nu,t_1)\d E_1\\ \nonumber
=\int p(I|E,t_0+\tau')N(E,\{\alpha\}_\nu,t_0+\tau')\d E.
\end{eqnarray}
Following the same steps that led to Equation~(\ref{eq:Nt2}) we find that
\begin{eqnarray}\label{eq:Nmarkov}
N(E,\{\alpha\}_\nu,t)=\int \d E_1 N(E_1,\{\alpha\}_\nu,t_1)p_c(E,t|E_1,t_1) \\ \nonumber
=\int \d E_0 N(E_0,\{\alpha\}_\nu,t_0)\int \d E_1p_c(E,t|E_1,t_1)p_c(E_1,t_1|E_0,t_0).
\end{eqnarray}
Hence, by analogy with Equation~(\ref{eq:Nt2}) one can define a transition probability $p_{c,2}(E,t|E_0,t_0)=\int \d E_1p_c(E,t|E_1,t_1)p_c(E_1,t_1|E_0,t_0)$.

The above result can be easily generalized to a serial sequence of $\kappa$ time-steps $t_0< t_1<,...,< t_{\kappa-1}\le t_\kappa$, with $t_\kappa=t_0+\tau'$. 
In doing this we find that the transition probability between $t_0$ and $t_0+\tau'$ can be expressed as a Markov chain (e.g. Krapivsky et al. 2010)
\begin{eqnarray}\label{eq:markov}
p_{c,\kappa}(E,t|E_0,t_0)=\int \d E_1 p_c(E_1,t_1|E_0,t_0)\int \d E_2 p_c(E_2,t_2|E_1,t_1)...\\ \nonumber
\times\int \d E_{\kappa-1}p_c(E,t|E_{\kappa-1},t_{\kappa-1})p_c(E_{\kappa-1},t_{\kappa-1}|E_{\kappa-2},t_{\kappa-2}).
\end{eqnarray}
We shall say that a Markov chain is {\it transitive} when the transition probability between the states $t_0$ and $t_0+\tau'$ is independent of the number of intermediate steps, that is if $p_c=p_{c,1}=...=p_{c,\kappa}$ for any value of $\kappa$.

As discussed in \S\ref{sec:timedf}, transition probabilities can be approached by Gaussian functions in potentials that evolve in a linear regime. It is straightforward to show that the mean ($\Delta \tilde C^j$) and the variance ($\tilde D_c^j$) of the Gaussian resulting from the integral over $\int \d E_j$ in~(\ref{eq:markov}) can be written as
\begin{eqnarray}
\label{eq:cc2}
\Delta \tilde C^j&=&\Delta\tilde C(E_{j+1},t_{j+1},E_{j,}t_{j})+\Delta\tilde C(E_{j},t_{j},E_{j-1},t_{j-1}) \bigg(\frac{R_{j}}{R_{j+1}}\bigg)^2,
%\\ \nonumber 
%&=&\tilde C(E_{j+1},t_{j+1})-\tilde C(E_{j-1},t_{j-1})\bigg(\frac{R_{j-1}}{R_{j+1}}\bigg)^2,
\end{eqnarray}
 and
\begin{eqnarray}
\label{eq:dc2}
\tilde D_c^j&=& \tilde D_c(E_{j+1},t_{j+1},E_j,t_j)+\tilde D_c(E_{j},t_{j},E_{j-1},t_{j-1})\bigg(\frac{R_j}{R_{j+1}}\bigg)^4.
%\\ \nonumber
%&=&\tilde D(E_{j+1},t_{j+1})+2\tilde D(E_j,t_j)\bigg(\frac{R_j}{R_{j+1}}\bigg)^4+ \tilde D(E_{j-1},t_{j-1})\bigg(\frac{R_{j-1}}{R_{j+1}}\bigg)^4.
\end{eqnarray}
Applying Equations~(\ref{eq:cc2}) and~(\ref{eq:dc2}) recursively from $j=\kappa-1$ to $j=1$, denoting $R=R(t_0+\tau')$ and $R_0=R(t_0)$, and using Equations~(\ref{eq:cc}) and~(\ref{eq:dc}) yields
\begin{eqnarray}
\label{eq:cc3}
\Delta \tilde C(E,t,E_0,t_0)= \tilde C(E,t)-\tilde C(E_0,t_0)\bigg(\frac{R_0}{R}\bigg)^2,
\end{eqnarray}
 and
\begin{eqnarray}\label{eq:dc3}
\tilde D_c(E,t,E_0,t_0)&=& \tilde D(E,t)+\tilde D(E_0,t_0)\bigg(\frac{R_0}{R}\bigg)^4+2\sum_{j=1}^{\kappa-1}\tilde D(E_j,t_j)\bigg(\frac{R_j}{R}\bigg)^4\\ \nonumber
&\approx& \frac{2}{\tau'}\int_{t_0}^{t_0+\tau'} \tilde D(E,x)\bigg[\frac{R(x)}{R}\bigg]^4\d x,
\end{eqnarray}
where the last step follows from changing $\sum_j\rightarrow \frac{1}{\tau'}\int \d t$ and applying the extended trapezoidal rule for $\kappa\gg 1$. Equations~(\ref{eq:cc3}) and~(\ref{eq:dc3}) respectively show that the drift coefficient $\Delta \tilde C$ retains no memory of the evolutionary path of the system, while $\tilde D_c$ is related to a time-average of the diffusion coefficient over the interval $t_0,t_0+\tau'$. 

Intuitively, we can see that the probability $p_c$ becomes transitive if the initial and final states do not deviate strongly from the time-averaged state of the system. This is clearly the case for time intervals $\tau'\ll (\dot \mu/\mu_0)^{-1}$, for which $(b-a)^{-1}\int_a^bf(t)\d t\approx [f(b)+f(a)]/2$, such that Equation~(\ref{eq:dc3}) reduces to~(\ref{eq:dc}). Therefore, we find that in potentials that evolve in a linear regime, where the time scale $(\dot \mu/\mu_0)^{-1}$ is very long compared with the orbital period of the particles, the transition (or ``jump'') between $t_0$ and $t_0+\tau'$ given by Equation~(\ref{eq:Nt2}) is transitive, i.e. it neither depends on the manner in which the intermediate evolution proceeds nor on the length of $\tau'$. It also becomes clear that in order to preserve the transitivity of Equation~(\ref{eq:Nt2}) in potentials that vary rapidly with time the length of the jumps must be decreased such that the condition $\tau'\ll (\dot \mu/\mu_0)^{-1}$ is everywhere and at all times satisfied.

%Given that $\tilde D\sim (\dot R/R)^2\sim (\dot \mu/\mu)^2$ and that $(b-a)^{-1}\int_a^bf(t)\d t\approx [f(b)+f(a)]/2$ for $a\approx b$, comparison between Equations~(\ref{eq:dc3}) and~(\ref{eq:dc}) indicates that $p_c$ behaves as a transitive function during time intervals $\tau'\lesssim (\dot \mu/\mu_0)^{-1}$.  This approximation is accurate in potentials that evolve in a linear regime, where the time scale $(\dot \mu/\mu_0)^{-1}$ is very long compared with the orbital periods.

%Below it is shown that the Fokker-Planck equations provide valuable insight into the behaviour of gravitating systems subject to time-dependent, long-range forces. 

\end{document}